\documentclass[a4paper,fleqn,usenatbib]{mnras}

\usepackage{newtxtext,newtxmath}
\usepackage[T1]{fontenc}
\usepackage{ae,aecompl}

\usepackage{graphicx}	% Including figure files
\usepackage{amsmath}	% Advanced maths commands
\title[Recurrent Nova T CrB]{The $B$ \& $V$ Light Curves for Recurrent Nova T CrB From 1842--2022, the Unique Pre- and Post-Eruption High-States, the Complex Period Changes, and the Upcoming Eruption in 2025.5$\pm$1.3}

\author[B. E. Schaefer]{
Bradley E. Schaefer$^{1}$\thanks{E-mail: schaefer@lsu.edu},
\\
$^{1}$Department of Physics and Astronomy, Louisiana State University, Baton Rouge, Louisiana, 70820, USA\\
}

\pubyear{2021}

\begin{document}
\label{firstpage}
\pagerange{\pageref{firstpage}--\pageref{lastpage}}
\maketitle

\begin{abstract}

T CrB is one of the most-famous and brightest novae known, and is a recurrent nova with prior eruptions in 1866 and 1946 that peak at $V$=2.0.  I have constructed light curves spanning 1842--2022 with 213,730 magnitudes, where the $B$ and $V$ magnitudes are fully corrected to the Johnson system.  These light curves first reveal a unique complex high-state (with 20$\times$ higher accretion rate than the normal low-state) stretching from -10 to +9 years after eruption, punctuated with a deep pre-eruption dip (apparently from dust formation in a slow mass ejection) and a unique enigmatic secondary eruption (with 10 per cent of the energy of the primary eruption), with the light curves identical for the 1866 and 1946 eruptions.  Starting in 2015, T CrB entered the high-state, like in 1936, so a third eruption in upcoming years has been widely anticipated.  With the pre-1946 light curve as a template, I predict a date of 2025.5$\pm$1.3 for the upcoming eruption, with the primary uncertainty arising from a possible lengthening of the pre-eruption high-state.  I use the large-amplitude ellipsoidal modulation to track the orbital phase of the binary from 1867--2022.  I measure that the orbital period  increased abruptly by $+$0.185$\pm$0.056 days across the 1946 eruption, the 1947--2022 years had a steady period decrease of ($-$8.9$\pm$1.6)$\times$10$^{-6}$ days-per-day, and the 1867--1946 years had a steady period change consistent with zero, at ($+$1.75$\pm$4.5)$\times$10$^{-6}$ days-per-day.  These large period changes cannot be explained by any published mechanism.
 
\end{abstract}


%%%%%%%%%%%%%%%%%%%%%%%%%%%%%%%%%%%%%%%%%%%%%%%%%%

%%%%%%%%%%%%%%%%% BODY OF PAPER %%%%%%%%%%%%%%%%%%

\section{INTRODUCTION}

T Coronae Borealis (T CrB) is a famous recurrent novae (RN), with very fast classical nova eruptions in 1866 and 1946 (Payne-Gaposchkin 1964).  T CrB in 1866 was the first well-observed nova and the first with spectroscopy (Huggins 1866).  T CrB peaks at 2.0 mag, making its 1946 eruption the brightest nova event from 1943--2022.  In quiescence, T CrB is by far the brightest of all known novae, with an average quiescent magnitude of 9.8 mag, and this allows for effectively continuous coverage of its light curves from 1866 to present, and with an average coverage of once-every-6-hours (even through its yearly solar conjunction) ever since 1946.  T CrB has the companion star to its white dwarf as a red giant star, M4 {\rm III}, which dominates the optical and infrared spectrum, all with an orbital period of 227.5687$\pm$0.0099 days (Kenyon \& Garcia 1986; Leibowitz, Ofek, \& Mattei 1997; Fekel et al. 2000).  The year-to-year light curve is dominated by ellipsoidal modulations at half the orbital period with typical full-amplitude of 0.3 mag, although rapid flickering at the 0.3 mag level is ubiquitous.

T CrB is also unique for having two separate eruptions, spaced half-a-year apart with an intervening  interval of 80 days stably at the pre-eruption brightness level.  The main eruption has a fast rise, a peak at $V$=2.0 mag, and a duration of 6 days within 3 mags of the peak, while the secondary event has a fast rise, a peak at $V$=8.0 mag, and a FWHM duration of 90 days.  The second-eruptions contain a substantial fraction of the radiative energy of the primary-eruptions.  This double-event was seen identically in 1866 and 1946.  These unprecedented second-eruptions have had little recognition in the literature, for which I am aware of no plausible explanation.  So here we have a highly-energetic new mode of nova eruptions that provides a mystery, as a challenge to theorists.

T CrB also has a unique and complex set of high-states, lasting roughly from $T_{\rm eruption}$-10 to $T_{\rm eruption}$+9 years, with year-long transitions between the normal quiescent low-state and the high-state (Schaefer 2014).  The high-state is prominent in blue light, with amplitude of 1.4 mag (Schaefer 2014), while the spectrum is dominated by the addition of high-ionization emission lines (Payne-Gaposchkin 1964). This high-state has an equal radiative energy as does the primary-eruption.  The post-eruption high-state light curve is identical between the 1866 and 1946 eruptions.  The recognition of the pre-eruption dip (the fast and deep dimming of T CrB in the months before the main eruption) was by L. Peltier in 1945, and he correctly interpreted this as a sign of an imminent eruption  (Peltier 1945).  The existence and details of the high-state was first recognized by Schaefer (2014), with this only being possible by my construction then of light curves in $B$ and $V$ with 102,000 magnitudes from 1855 to 2013, where no one had previously constructed an adequately long and well-calibrated data set.  I know of no attempts to explain this complex high-state.  I am struck by the difficulty to explain how the light curve rise-and-falls for the pre-eruption high-state can {\it anticipate} the upcoming classic nova event.  So we have another T CrB mystery, one that dominates the energetics, as a challenge for theorists.

RNe are classical novae which have recurrence time-scales ($\tau_{rec}$) of shorter than 100 years. Only 10 systems are now known in our Milky Way Galaxy with multiple discovered eruptions separated by less than 100 years (Schaefer 2010), while one other system has a recurrence time scale of 40--50 years (Schaefer et al. 2022).  The last two eruptions of T CrB were separated by 80 years, so the simplistic idea is that the next eruption will be around the year 1946+80, or 2026.  This schematic prediction has been common knowledge at least since my undergraduate days in the 1970s.  The expected accuracy of this calculation is poor, since the observed ratio of longest-to-shortest recurrence time-scales is a factor of 2.1$\times$ for U Sco, 3.7$\times$ for T Pyx, and 2.9$\times$ for RS Oph.  With a discovery of the anticipatory pre-eruption high-state and dip, a new possibility opened up for a means to get an accurate prediction of the upcoming eruption (Schaefer 2014).  Then, in the year 2015, the American Association of Variable Star Observers (AAVSO) $B$ light curve had a sharp transition to a high-state, with the morphology of the $B$ and $V$ light curves being closely similar to that around 1938.  The 2015 transition to a high-state was first recognized photometrically and spectroscopically by Munari, Dallaporta, \& Cherini (2016), who called attention to this as being similar to the transition in 1936.  Assuming that the light curve of the pre-eruption high-state is similar from eruption-to-eruption, Schaefer (2019) predicted the next eruption for 2023.6$\pm$1.0.  In a follow-up with a subset of my data, Luna et al. (2020) predict that the eruption will be 2023--2026.  Currently, there is a widespread anticipation that T CrB will erupt soon.

\section{Light Curve from 1842--2022}

A primary purpose of this paper is to construct a complete light curve, in the modern Johnson $B$ and $V$ systems, from 1842--2022.  Table 1 contains a listing of the observers and their details.  Individual magnitudes are explicitly listed in Table 2 for the visual observations in Section 2.1, in Table 3 for the photographic magnitudes from archival plates, in Table 4 for the photoelectric and CCD observations from the literature in Section 2.3, and in Table 5 for the collected observations from amateur observers worldwide in Section 2.4.  Figure 1 shows the overall plot of the 1842--2022 $B$ and $V$ light curves.

\begin{table*}
	\centering
	\caption{T CrB photometry data sources}
	\begin{tabular}{llllll} 
		\hline
		Observer & Years & Start (JD)  &  Band   &   Count   &   Source  \\
		\hline	
J. F. W. Herschel	&	1842	&	2393996	&	{\it Vis.}$\rightarrow${\it V}	&	1	&	Schaefer (2013)	\\
F. Argelander	&	1855--1856	&	2398721	&	{\it Vis.}$\rightarrow${\it V}	&	2	&	{\it Bonner Durchmusterung}, c.f. Schoenfeld (1875)	\\
J. F. J. Schmidt	&	1866--1879	&	2402734	&	{\it Vis.}$\rightarrow${\it V}	&	936	&	Schmidt (1877; 1879)	\\
J. Birmingham	&	1866	&	2402734	&	{\it Vis.}$\rightarrow${\it V}	&	1	&	Birmingham (1866)	\\
Courbebaisse	&	1866	&	2402734	&	{\it Vis.}$\rightarrow${\it V}	&	1	&	Stone (1866)	\\
S. C. Chandler	&	1866	&	2402736	&	{\it Vis.}$\rightarrow${\it V}	&	17	&	Gould (1866a; 1866b)	\\
C. H. Davis	&	1866	&	2402736	&	{\it Vis.}$\rightarrow${\it V}	&	15	&	Davis (1866)	\\
J. Baxendell	&	1866--1869	&	2402737	&	{\it Vis.}$\rightarrow${\it V}	&	101	&	Baxendell (1866--1869)	\\
J. Carpenter	&	1866	&	2402739	&	{\it Vis.}$\rightarrow${\it V}	&	2	&	Stone (1866)	\\
F. Bird	&	1866	&	2402740	&	{\it Vis.}$\rightarrow${\it V}	&	18	&	Bird (1866)	\\
W. R. Dawes	&	1866	&	2402740	&	{\it Vis.}$\rightarrow${\it V}	&	5	&	Dawes (1866a; 1866b)	\\
E. J. Stone	&	1866	&	2402740	&	{\it Vis.}$\rightarrow${\it V}	&	7	&	Stone (1866)	\\
E. Schoenfeld	&	1866--1875	&	2402744	&	{\it Vis.}$\rightarrow${\it V}	&	468	&	Campbell (1920); Valentiner (1900)	\\
C. Behrmann	&	1866	&	2402746	&	{\it Vis.}$\rightarrow${\it V}	&	5	&	Behrmann (1866)	\\
T. W. Backhouse	&	1866--1916	&	2402751	&	{\it Vis.}$\rightarrow${\it V}	&	429	&	Backhouse (1905; 1916)	\\
A. Kreuger	&	1866--1867	&	2402886	&	{\it Vis.}$\rightarrow${\it V}	&	12	&	Heis \& Krueger (1903)	\\
H. M. Parkhurst	&	1884, 1892	&	2409375	&	{\it Vis.}$\rightarrow${\it V}	&	3	&	Parkhurst (1890)	\\
Harvard 	&	1890--1962	&	2411572	&	{\it B}	&	896	&	Harvard plates (this paper)	\\
T. E. Espin	&	1893, 1899	&	2412563	&	{\it Vis.}$\rightarrow${\it V}	&	2	&	Espin (1893; 1900)	\\
J. Holetschek	&	1896--1909	&	2413869	&	{\it Vis.}$\rightarrow${\it V}	&	14	&	Holotschek (1907, 1912)	\\
E. E. Barnard	&	1906	&	2417451	&	{\it Vis.}$\rightarrow${\it V}	&	2	&	Barnard (1907)	\\
E. Zinner	&	1913--1935	&	2419975	&	{\it Vis.}$\rightarrow${\it V}	&	40	&	Ferrari (1935)	\\
W. H. Steavenson	&	1925--1947	&	2424360	&	{\it Vis.}$\rightarrow${\it V}	&	82	&	Steavenson (1926--1948)	\\
K. Ferrari	&	1929--1935	&	2425687	&	{\it B}	&	82	&	Bamberg plates (Ferrari 1935)	\\
Bamberg	&	1932--1939	&	2426868	&	{\it B}	&	16	&	Bamberg plates (this paper)	\\
S. Bohme	&	1935--1938	&	2427932	&	{\it B}	&	57	&	Bamburg plates (Bohme 1938)	\\
K. Himpel	&	1936--1938	&	2428248	&	{\it Vis.}$\rightarrow${\it V}	&	153	&	Himpel (1938a; 1938b)	\\
Sonneberg	&	1936--2000	&	2428422	&	{\it B}	&	692	&	Sonneberg plates (this paper)	\\
1515 AID observers	&	1939--2022	&	2429382	&	{\it Vis.}$\rightarrow${\it V}	&	114203	&	AAVSO AID$^a$	\\
A. DeutschW. W. Morgan	&	1946	&	2431860	&	{\it Vis.}$\rightarrow${\it V}	&	17	&	Morgan \& Deutsch (1947)	\\
N. F. H. Knight	&	1946	&	2431860	&	{\it Vis.}$\rightarrow${\it V}	&	2	&	Knight (1946)	\\
E. Pettit	&	1946--1950	&	2431861	&	{\it Vis.}$\rightarrow${\it V}	&	175	&	Pettit (1946; 1950)	\\
J. Ashbrook	&	1946	&	2431861	&	{\it Vis.}$\rightarrow${\it V}	&	54	&	Ashbrook (1946)	\\
K. C. Gordon \& G. E. Kron	&	1946	&	2431861	&	{\it B}	&	14	&	Gordon \& Kron (1979)	\\
K. C. Gordon \& G. E. Kron	&	1946	&	2431862	&	{\it V}	&	10	&	Gordon \& Kron (1979)	\\
M. Ch. Bertaud	&	1946--1947	&	2431865	&	{\it Vis.}$\rightarrow${\it V}	&	58	&	Bertaud (1947)	\\
R. Weber	&	1946--1961	&	2431934	&	{\it B}	&	124	&	Weber (1961)	\\
113 AID observers	&	1973--2022	&	2441832	&	{\it V}	&	7368	&	AAVSO AID$^a$	\\
H. C. Lines et al.	&	1981--1983	&	2444715	&	{\it V}	&	83	&	Lines et al. (1988)	\\
H. C. Lines et al.	&	1982--1983	&	2445141	&	{\it B}	&	57	&	Lines et al. (1988)	\\
D. Raikova \& A. Antov	&	1985	&	2446200	&	{\it B}	&	16	&	Raikova \& Antov (1986)	\\
D. Raikova \& A. Antov	&	1985	&	2446200	&	{\it V}	&	16	&	Raikova \& Antov (1986)	\\
L. Hric et al.	&	1990--1997	&	2447969	&	{\it B}	&	88	&	Hric et al. (1998)	\\
L. Hric et al.	&	1990--1997	&	2447969	&	{\it V}	&	87	&	Hric et al. (1998)	\\
R. Zamanov et al.	&	1991--1999	&	2448321	&	{\it B}	&	95	&	Zamanov \& Zamanova (1997); Zamanov et al. (2004)	\\
R. Zamanov et al.	&	1991--1999	&	2448321	&	{\it V}	&	95	&	Zamanov \& Zamanova (1997); Zamanov et al. (2004)	\\
ASAS	&	2003--2009	&	2452689	&	{\it V}	&	236	&	Pojmanski (1997)$^b$	\\
52 AID observers	&	2004--2022	&	2453075	&	{\it B}	&	3937	&	AAVSO AID$^a$	\\
U. Munari et al.	&	2006--2015	&	2453867	&	{\it B}	&	204	&	Munari et al. (2016)	\\
U. Munari et al.	&	2006--2015	&	2453867	&	{\it V}	&	205	&	Munari et al. (2016)	\\
APASS	&	2012	&	2455989	&	{\it B}	&	10	&	AAVSO APASS$^c$	\\
APASS	&	2012	&	2455989	&	{\it V}	&	10	&	AAVSO APASS$^c$	\\
{\it TESS} Sector 24	&	2020	&	2458955	&	{\it TESS}	&	16119	&	{\it TESS} SPOC from MAST$^d$	\\
{\it TESS} Sector 25	&	2020	&	2458983	&	{\it TESS}	&	17116	&	{\it TESS} SPOC from MAST$^d$	\\
{\it TESS} Sector 51	&	2022	&	2459698	&	{\it TESS}	&	49272	&	{\it TESS} SPOC from MAST$^d$	\\
		\hline
	\end{tabular}
	
\begin{flushleft}	
$^a$AAVSO International Database, \url{https://www.aavso.org/data-download}\\
$^b$All Sky Automated Survey, \url{http://www.astrouw.edu.pl/asas/?page=aasc}\\
$^c$AAVSO Photometric All-Sky Survey, \url{https://www.aavso.org/download-apass-data}\\
$^d$Barbara A. Mikulski Archive for Space Telescopes, \url{https://mast.stsci.edu/portal/Mashup/Clients/Mast/Portal.html} \\
\end{flushleft}

\end{table*}

\begin{figure*}
	\includegraphics[width=2.08\columnwidth]{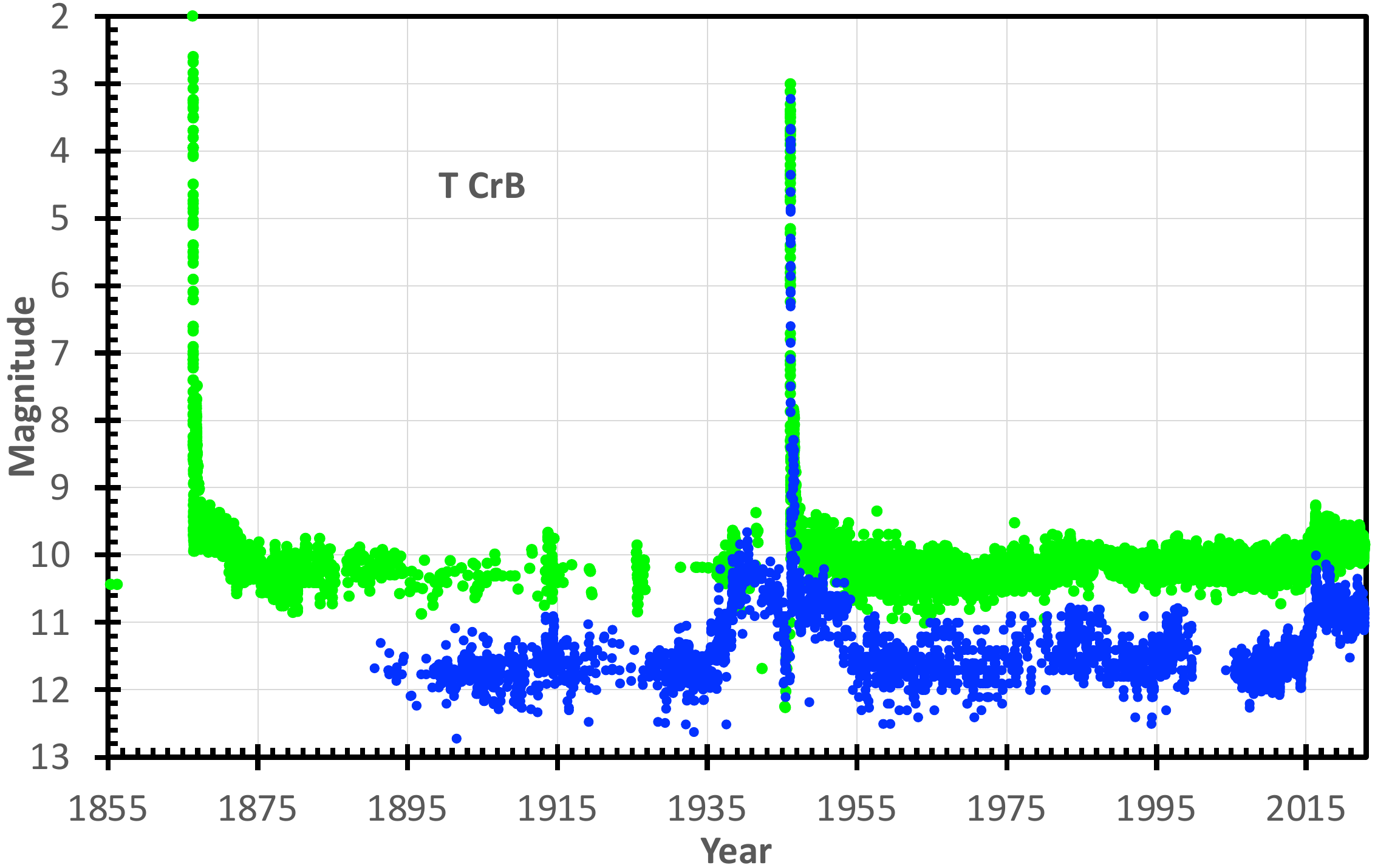}
    \caption{Light curve for T CrB.  All $B$- and $V$-band observations are shown by the blue and green circles respectively.  We see the two eruptions in 1866 and 1946, reaching up to $V$=2.0.  The scatter seen in any year is entirely from the usual flickering and the 113-day ellipsoidal modulation.  (The measurement error bars are comparable in size to the dots or smaller.)  We see that the $B-V$ varying with a typical value around 1.2 mag.  Most importantly, this graph, or its equivalent, is the only way to see complex high-state from 1866--1875, 1936--1955, and 2015--present.  These high-states are the same from eruption-to-eruption.  These high-states consist mostly of blue-light, with them prominent in $B$ and only noticeable in $V$.  The pre-eruption high-state starts nearly 10 years before the eruption, so the high-state starting around 2015 implies an upcoming third eruption around the year 2025.  The pre-eruption dip will provide an immediate notice of a few months before the upcoming eruption.  The secondary eruption in 1866 and 1946 is lost with the primary eruption in this compressed horizontal scale.  The high-state has the same total energy as the the classical nova eruptions, while the T CrB high-state is unique.}  
\end{figure*}

\subsection{Visual Observations}

Visual magnitude measures are the traditional method by which an observer directly compares the brightness of the target star to that of a nearby comparison star.  With a sequence of comparison stars, the observer place the brightness as some fraction between the nearest comparison stars that are just-fainter and just-brighter.  With a knowledge of the adopted magnitudes for the sequence stars, the target's magnitude is just the fractional difference in these comparison star magnitudes.  With moderate practice plus good nearby comparison stars, a 1-sigma photometric accuracy of 0.20 mag is easily obtained (Stanton 1999).  The most experienced observers can consistently get 0.10 or even 0.07 mag accuracy.  

Visual magnitudes constitute 89 per cent of the magnitudes in Fig. 1.  Most of the science in this paper is possible only because of the vast numbers and complete time coverage of the visual measures.  I have collected 116,844 visual magnitudes for T CrB, as itemized in Table 1.  The bulk of the visual magnitudes are from the AAVSO International Database (AID), all from 1939 and later, and are considered separately in Section 2.4.  Here, I will only look at the 2621 visual magnitudes not in the AID (see Table 2).  The sources are mainly in widely scattered published papers (mostly with amateur observers), and in manuscripts now stored in various archives in Germany, England, and the US.  The entire $V$ light curve from 1842--1939 and the primary measures of both eruption light curves are derived from these visual observations. 

All of the visual magnitudes should be converted to $V$.  Visual magnitudes have a similar spectral sensitivity as Johnson $V$ magnitudes, with this by intentional construction.  This means that visual and $V$ magnitudes are similar, yet with significant systematic offsets.  This conversion is based on the massive study of Stanton (1999), involving 63 observers of a wide range of ages, experiences, and equipment, including three colour blind people.  The conversion is simply 
\begin{equation}
V = m_{\rm vis} - 0.210(B-V),
\end{equation}
where $m_{\rm vis}$ is the visual magnitude as reported by a visual observer, the target colour index is the usual $B-V$, and $V$ is the Johnson $V$ measure.  For the case of T CrB, the average colour in quiescence is $B-V$=1.22 mag (Bruch \& Engel 1994) with only small variations. During eruption, the $B-V$ can be found from the light curves in Schaefer (2010).  We have no direct measure of the colour at peak light, but this is $(B-V)_{\rm peak}$=$+$0.11 mag for apparently all classical novae (Schaefer 2022c), while T CrB has near-zero extinction due to its high Galactic latitude and small distance.  So we now have a method for converting from the observed $m_{\rm vis}$ to $V$.

The observers almost always reports their magnitudes where they have already worked out the visual magnitude in comparison with their sequence.  An ubiquitous trouble is that the observers' adopted magnitudes for their comparison sequences are skewed from the Johnson $V$ magnitudes, often by simple rounding of the sequence star magnitudes to the nearest tenth, and often by systematic errors that can get up to a magnitude is size.  Another ubiquitous trouble is that the adopted magnitudes and the catalogued magnitudes are never in the visual magnitude system.  The general solution is presented in Johnson et al. (2014).  The first step is to look up the Johnson $B$ and $V$ magnitudes for all the sequence stars, and use Eq. 1 to calculate $m_{\rm vis}$ for each star.  The second step is to calculate the fraction for which the reported magnitude is between the adopted magnitudes for the two stars just-brighter and just-fainter than the target, with this backtracking the observer's calculation when they reported the magnitude.  The third step is to take this fraction to be the observed fraction that the target is between the visual magnitudes of the two comparison stars.  The fourth step is to equate the two equations for the fraction, and solve for the visual magnitude of the target.  The fifth step is to convert the target visual magnitude to its $V$ magnitude.  The resulting equation is 
\begin{equation}
V=m_b + [(m_f-m_b)(\mu - \mu_b)/(\mu_f - \mu_b)] -0.21\times(B-V).
\end{equation} 
Here, $m_b$ and $m_f$ are the visual magnitudes of the sequence stars just brighter-than and just fainter-than the target, with these being derived from catalogue $B$ and $V$ for the sequence stars, along with an application of Eq. 1.  The sequence has adopted the magnitudes $\mu_b$ and $\mu_f$ for these stars.  Based on this sequence, the observer makes a judgment on the relative brightness of the target and reports a magnitude $\mu$.  The $B-V$ is the colour of the target star.  Fortunately, essentially all published reports of visual observations explicitly identify their comparison stars (either by a chart or by coordinates), and they explicitly state their adopted magnitudes for each sequence star.  The end result is a confident conversion from visual magnitudes to the Johnson $V$ magnitude system.

The first observation in my light curve is that of Sir John Herschel on 1842 June 9, as part of his visual survey of the entire sky.  He noted a star near the position of T CrB, and when the nova erupted in 1866, he claimed that T CrB was near 6 mag in 1842 (Herschel 1866).  If correct, then T CrB had an eruption in 1842.  This identification has been questioned by McLaughlin (1939) due to the published chart showing the star at the position of a nearby star (HD 144287) with a stable $V$=7.06.  But such a star could not have been seen with Herschel's nominal naked-eye charting.  In this ambiguous situation, I have made an exhaustive examination of Herschel's correspondence, notebooks, diary, and papers (Schaefer 2013).  I found a chart made by Herschel with star positions guaranteed by pinpricking his original chart, sent in a letter to W. Huggins (dated 1866 May 19), and see that the position of the star is that of HD 144287 (and inconsistent with T CrB).  Further, I found Herschel's notes telling of his use of an opera glass so as to get to 7 mag.  So there was no 1842 eruption, and Herschel's observation is actually a limit on T CrB of $V$$>$7.06.

The great Leslie Peltier started regularly monitoring T CrB in 1920, before the existence of RN was known, on a general idea that it would have another nova event.  For composite light curve purposes, I cannot use any of his 236 step estimates from 1920--1933 (which I have only found in letters saved in the AAVSO files), nor his 249 magnitude measures from 1933--1946.20.  One basis for this exclusion is that 180 out of 249 of his magnitudes are reported as exactly 9.8 mag (233 are 9.7--9.9 mag), with T CrB not behaving as such.  Peltier's magnitudes disagree by up to a magnitude from everyone else in the world.  And Peltier's comparison star sequence (given in a private letter to Leon Campbell dated 3 June 1923) has differences from the modern $V$ that range from -0.02 to -1.68 mag for stars over the narrow magnitude range of 9.2--9.76.  Seventeen days after the end of this series, Peltier discovered the pre-eruption dip, realized that it meant an eruption was imminent, and announced it to the world (Peltier 1945).  Peltier's long vigil on T CrB (`We had been friends for many years; on thousands of nights I had watched over it as it slept') just barely missed the discovery of the 1946 eruption,  and then `There is no warmth between us anymore' (Peltier 1965).

In the end, I have 2621 visual magnitude measures that have been converted to the Johnson $V$ system.  As the original sources for the magnitudes did not provide error estimates, $\pm$0.2 mags is adopted.

\begin{table}
	\centering
	\caption{T CrB visual photometry (full table with 2621 lines available on-line as Supplementary Material)}
	\begin{tabular}{lllll} 
		\hline
		Julian Date & Year & Band  &  Magnitude  &   Source   \\
		\hline
2393996.5	&	1842.441	&	{\it Vis.}$\rightarrow${\it V}	&	$>$7.09	$\pm$	0.20	&	J. Herschel	\\
2398721.5	&	1855.379	&	{\it Vis.}$\rightarrow${\it V}	&	10.44	$\pm$	0.20	&	Argelander	\\
2399039.6	&	1856.248	&	{\it Vis.}$\rightarrow${\it V}	&	10.44	$\pm$	0.20	&	Argelander	\\
2402734.37	&	1866.364	&	{\it Vis.}$\rightarrow${\it V}	&	$>$5	$\pm$	0.20	&	Schmidt	\\
2402734.486	&	1866.363	&	{\it Vis.}$\rightarrow${\it V}	&	2.00	$\pm$	0.20	&	Birmingham	\\
...	&		&				&		&		\\
2433011.001	&	1949.257	&	{\it Vis.}$\rightarrow${\it V}	&	9.67	$\pm$	0.20	&	Pettit	\\
2433061.705	&	1949.395	&	{\it Vis.}$\rightarrow${\it V}	&	9.71	$\pm$	0.20	&	Pettit	\\
2433119.698	&	1949.554	&	{\it Vis.}$\rightarrow${\it V}	&	9.76	$\pm$	0.20	&	Pettit	\\
2433186.649	&	1949.738	&	{\it Vis.}$\rightarrow${\it V}	&	9.73	$\pm$	0.20	&	Pettit	\\
2433313.040	&	1950.084	&	{\it Vis.}$\rightarrow${\it V}	&	9.67	$\pm$	0.20	&	Pettit	\\
		\hline
	\end{tabular}

\end{table}

\subsection{Photographic Magnitudes}

Before the advent of modern electronic detectors, the only way to get a light curve in some colour other than the visual-band was to use photography.  In practice, a number of observatories collected large numbers of sky photographs, where the emulsion was on one side of a clear glass plate, so the stars appeared as negative images on these plates.  The brightness of the star is effectively always measured from the sharply-defined image radius, where the magnitude scale is always calibrated by the image radii of nearby stars of known brightness.  Before the 1970s, the spectral sensitivity of the emulsions were nearly always indistinguishable from that of the modern Johnson $B$ magnitude system, and does not vary significantly over time (Laycock et al. 2010).  Therefore, when the magnitudes for the calibrating comparison stars are in the Johnson $B$ magnitude system, the result magnitude will be accurately in the modern Johnson $B$ system.

T CrB is bright, and so is readily recorded on most archival plates.  The largest collection of archival astronomical plates is at the Harvard College Observatory (HCO), where I have measured 896 magnitudes from 1890--1962.  The Harvard plates are always the only source of $B$ magnitudes before the 1930s.  The second largest collection of archival astronomical plates is at Sonneberg Observatory, in Germany, where I have measured 692 plates from 1936--2000.  The Sonneberg plates are valuable as usually being the only source of $B$ magnitude starting in 1954 (with the notorious Menzel Gap at HCO) up until electronic detectors became common in the 1990s.  The plate collection at Bamberg Observatory, in Germany, has a good collection of plates that covers the 1930s.  Bohme (1938) and Ferrari (1935) have already published 139 $B$ magnitudes for T CrB, although I have had to convert their old magnitudes to modern $B$ magnitudes by detailed analysis of their stated comparison stars and their adopted sequences of magnitudes.  Further, I have found one previously-unused box of plates in the Bamberg archives, and this provides 16 more $B$ magnitudes from 1932--1939.  A further set of archival plates is presented in Weber (1961), with 124 magnitudes from 1946--1961, all from his private observatory.    Weber was using 103a-O emulsion (with $B$ spectral sensitivity) and I have had to convert to modern $B$ magnitudes by using his stated comparison stars and sequence.  Weber's light curve is valuable as having the best time coverage in $B$ for the secondary eruption in 1946.

My standard technique for measuring the magnitudes from plates are told in detail in Schaefer (2016a; 2016b).  Importantly, I have used comparison stars with $B$ magnitudes from APASS, with these chosen to have red colours similar to T CrB itself.  The $B$ magnitudes of T CrB for many individual plates have been measured 6 times by myself and others.  I have made trips to HCO in 2008, 2010, and 2013, making measures from 896 plates with many duplicate measures.  A. Pagnotta (College of Charlston) made 19 measures of plates, designed to test the reproducibility and accuracy of the magnitudes.  Shapley (1933) reports on 342 measures (not counting limits) made by the highly experienced H. Leavitt (Harvard), which I have had to convert to modern $B$ magnitudes as derived from their comparison star sequence.  Finally, the large-scale programme Digital Access to a Sky Century $@$ Harvard (DASCH), with J. Grindlay (Harvard) as Principle Investigator, has scanned and digitized the majority of Harvard plates, and used those scans to run a sophisticated program to calculate the $B$ magnitudes of all stars on the plates (Tang et al. 2013).  DASCH reports 354 magnitudes.

This large-scale  study of T CrB provides yet another study of the measurement accuracy of the photographic photometry.  In all, these multiple-measures shows that the six independent magnitudes from the four sources all agree with each other closely.  For the three non-DASCH sources, the average difference in magnitudes is 0.009 mag, which shows that no one source is systematically making measures bright or dim.  The comparison between all the non-DASCH sources shows the same RMS of the differences, so the errors for each individual source is similar and near $\pm$0.15 mag.  That is, these sources are accurate with an unbiased real measurement error of 0.15 mag, for the T CrB case.  The comparisons of the differences between the DASCH magnitudes and from the three other sources shows a consistently larger RMS, which corresponds to the real measurement error for DASCH of $\pm$0.35 mag for this one case of T CrB.  For the final combined magnitudes for each HCO plate, I used an average of all the individual measures, and I'll adopt an error bar of $\pm$0.15 mag.

In the end, I have 1867 $B$ magnitudes, mostly measured by myself from the plates in front of me.  These measures are important because they provide the only measure of the blue light from 1890--1982, including all the complex high-state and eruption phenomena from 1938--1955.  These magnitudes are in Table 3 and plotted in Fig. 1.

\begin{table}
	\centering
	\caption{T CrB $B$ light curve from photographic photometry (full table with 1867 lines available on-line as Supplementary Material)}
	\begin{tabular}{lllll} 
		\hline
		Julian Date & Year & Band  &  Magnitude  &   Source   \\
		\hline
2411572.599	&	1890.560	&	{\it B}	&	11.68	$\pm$	0.15	&	HCO (I1520)	\\
2411878.689	&	1891.398	&	{\it B}	&	11.30	$\pm$	0.15	&	HCO (I3634)	\\
2412250.715	&	1892.417	&	{\it B}	&	11.70	$\pm$	0.15	&	HCO (I6373)	\\
2412262.672	&	1892.449	&	{\it B}	&	11.71	$\pm$	0.15	&	HCO (I6455)	\\
2412287	&	1892.516	&	{\it B}	&	11.76	$\pm$	0.15	&	HCO (I)	\\
...	&		&				&		&		\\
2451055.482	&	1998.660	&	{\it B}	&	12.00	$\pm$	0.21	&	Sonneberg	\\
2451427.476	&	1999.678	&	{\it B}	&	11.80	$\pm$	0.21	&	Sonneberg	\\
2451428.478	&	1999.681	&	{\it B}	&	11.80	$\pm$	0.21	&	Sonneberg	\\
2451430.475	&	1999.686	&	{\it B}	&	11.60	$\pm$	0.21	&	Sonneberg	\\
2451661.563	&	2000.319	&	{\it B}	&	11.40	$\pm$	0.21	&	Sonneberg	\\
		\hline
	\end{tabular}

\end{table}

\subsection{Photoelectric Photometers and CCDs}

Electronic detectors have been available since the 1946 eruption for single-channel photoelectric photometers and since roughly the year 2000 for CCD photometry.  For photoelectric photometers, five publications have reported 561 $B$ or $V$ magnitudes from 1946--1999 (see Table 1), with these being useful to check the light curves from other sources.  For T CrB magnitudes with CCDs, I have only found three sources from professional astronomers with 665 magnitudes 2003--2015.  Further, the AID records 11,305 $B$ and $V$ measures from 1973--2022 from photoelectric photometers and CCDs.

As with all photometry where magnitudes from many different observers are being combined, we have to be careful that the magnitude systems are close to the Johnson systems for each observer.  The electronic detectors are linear in flux, so we need not worry about scale changes from bright to faint stars.  For the selected T CrB measures, the detectors have always used filters that produce a spectral sensitivity close to those of the Johnson $B$ and $V$ systems, so colour terms will be negligible.  The adopted $B$ and $V$ magnitudes for their primary comparison stars are always within the usual 0.03 mag of the values now listed in the SIMBAD database.  Indeed, the APASS magnitudes and calibrations are now serving as the standard system.  In all, there are no corrections applied to the photoelectric and CCD magnitudes, their systematic errors are negligibly small, and their measurement errors are typically $\sim$0.03 mag or smaller.  The good photometric precision from Poisson noise for the electronic detectors has no utility for most science questions concerning T CrB, because the star has intrinsic random flickering at the 0.3 mag level, so the sampling error always dominates over the measurement error.

The 1226 photoelectric and CCD magnitudes are in Table 4.

\begin{table}
	\centering
	\caption{T CrB light curve from photoelectric and CCD photometry (full table with 1226 lines available on-line as Supplementary Material)}
	\begin{tabular}{lllll} 
		\hline
		Julian Date & Year & Band  &  Magnitude  &   Source   \\
		\hline
2431861.961	&	1946.111	&	{\it B}	&	3.68	0.03	&	Gordon \& Kron	\\
2431862.021	&	1946.111	&	{\it B}	&	3.66	0.03	&	Gordon \& Kron	\\
2431862.063	&	1946.111	&	{\it V}	&	3.56	0.03	&	Gordon \& Kron	\\
2431863.925	&	1946.116	&	{\it V}	&	4.58	0.03	&	Gordon \& Kron	\\
2431863.951	&	1946.116	&	{\it B}	&	4.85	0.03	&	Gordon \& Kron	\\
...	&		&				&		&		\\
2457349.216	&	2015.891	&	{\it V}	&	9.809	0.015	&	Munari et al.	\\
2457369.676	&	2015.947	&	{\it B}	&	10.924	0.008	&	Munari et al.	\\
2457369.676	&	2015.947	&	{\it V}	&	9.821	0.012	&	Munari et al.	\\
2457376.666	&	2015.966	&	{\it B}	&	10.944	0.009	&	Munari et al.	\\
2457376.666	&	2015.966	&	{\it V}	&	9.797	0.015	&	Munari et al.	\\
		\hline
	\end{tabular}

\end{table}

\subsection{AAVSO International Database}

Amateur astronomers have been keeping long vigils on T CrB since 1866.  Most of the visual observations reported in Section 2.1 are by amateurs, all with a photometric accuracy that is the same as for professional observers.  Mostly starting with the 1946 eruption, the worldwide coverage by top-quality amateur observers has made T CrB one of the best observed of all variable stars.  From 1946 to 2022, the average is 4 visual measures per night, every night.  The amateurs started making well-calibrated photoelectric observations in 1973, and well-calibrated CCD observations around 2004.

This massive coverage makes the `amateur' contributions to be the most important out of all the sources for the T CrB light curve.  Little of this has been published in the professional literature.  The only way to access most of this critical data is through dusty notebooks in archives, and in the repositories of various variable star organizations around the globe.  Prominent archives are the variable star sections of the British Astronomical Association (BAA), the French Variable Star Observers Association (AFOEV), the Royal Astronomical Society of Canada, and even the Royal Astronomical Society of New Zealand.  Close to half of all observers and all magnitudes are from people based in the United States with an affiliation with the American Association of Variable Star Observers (AAVSO), with headquarters in Cambridge Massachusetts.  The AAVSO performs the invaluable service of collecting all magnitudes from every available source worldwide, and placing them uniformly in one publicly-available database, the AAVSO International Database (AID).

I have downloaded all the magnitudes in the AID up until October 2022.  I concentrate only on the $B$, $V$, and visual magnitude estimates.  I do not use any measures that are only limits on the magnitude, or for which the stated uncertainty is $>$0.30 mag.  Nor do I use any unfiltered CCD measures calibrated with $V$ comparison stars, the so-called $CV$ magnitudes, as these have colour corrections that are always unknown in particular and can be at the 0.10 mag level or larger.  I am left with 52 observers making 3937 $B$-band measures with CCDs from 2004 to present, 113 observers making 7368 $V$-band measures mostly with CCDs from 1973 to present, and 1515 observers making 114,203 visual observations from 1939 to present.

The photometric accuracy of the amateurs is always the same as for the professional observers, and both are more than accurate enough to answer the various questions raised in this paper.  For the CCD and photoelectric measures, the typical uncertainty is $\sim$0.03 mag, even though the Poisson error might be substantially smaller.  For the visual observations, the typical uncertainty is close to 0.20 mag (Stanton 1999).  When binned together, say in 0.01-year bins as I do, with an average of near 16 observations per bin, the formal photometric measurement uncertainty is around 0.05 mag.  Nevertheless, the total uncertainty is not from the measurement errors, but rather arises from the intrinsic random variability of T CrB itself.  That is, ordinary flickering of T CrB causes fast variations up to the 0.5-mag level, and this dwarfs the measurement errors from all sources, professional or amateur.  When T CrB is up-and-down by half-a-magnitude on all time-scales, it matters little whether there is some additional measurement error at the 0.05 mag level.  This means that the formal measurement errors have little meaning and no utility.  Rather, the uncertainty of the magnitude is dominated by the sampling errors.  So the utility of an eyeball estimate by an amateur with an uncertainty of $\pm$0.20 mag is essentially the same as the utility of a well-calibrated CCD measure by a professional with an uncertainty of 0.01 mag.  However, the utility of averaging 16 visual measures on 16 nights is substantially better (4$\times$ in this case) than any single measure with a photometric accuracy of 0.001 mag, with the reason being that the many-night-combinations have averaged over the ubiquitous flickering on T CrB.

The visual magnitudes need to be accurately converted to $V$ magnitudes.  The procedure is described in Section 2.1.  Critical for this procedure is to know the comparison stars used by the individual observers.  Fortunately, observers worldwide largely used the same charts and sequences, and these sequences had only insignificant changes over time after 1939.  For this, I have found charts and sequences as used by observers in the archives of the AAVSO, BAA, and elsewhere.  On this basis, I have converted all of the visual magnitudes to the modern Johnson $V$ magnitude system.

The important variability, for the purposes of this paper, are all on time-scales of weeks to months to years.  As such, I have binned up the AID $B$ and $V$ light curves to 0.01-year bins.  The exception is during the months of the 1866 and 1946 eruptions, where the fast variations are better represented with much smaller bin sizes.  The result is a binned light curve with 819 $B$ measures, 1137 $V$ measures, and 7127 visual measures converted to $V$ (see Table 5 and Fig. 1).  In the source column, the parentheses states either the AAVSO observer ID or the number of magnitudes averaged together.

\begin{table}
	\centering
	\caption{T CrB $B$ and $V$ light curve from photometry in AAVSO International Database, mostly binned to 0.01-years (full table with 9083 lines available on-line as Supplementary Material)}
	\begin{tabular}{lllll} 
		\hline
		Julian Date & Year & Band  &  Magnitude  &   Source   \\
		\hline
2429382.4000	&	1939.3221	&	{\it Vis.}->{\it V}	&	10.207	$\pm$	0.093	&	AID (3)	\\
2429403.4000	&	1939.3796	&	{\it Vis.}->{\it V}	&	10.490	$\pm$	0.150	&	AID (LT)	\\
2429423.4000	&	1939.4344	&	{\it Vis.}->{\it V}	&	10.660	$\pm$	0.106	&	AID (2)	\\
2429436.4000	&	1939.4700	&	{\it Vis.}->{\it V}	&	10.490	$\pm$	0.150	&	AID (LT)	\\
2429452.9000	&	1939.5151	&	{\it Vis.}->{\it V}	&	10.575	$\pm$	0.106	&	AID (2)	\\
...	&		&				&		&		\\
2459855.8873	&	2022.7540	&	{\it Vis.}->{\it V}	&	9.804	$\pm$	0.040	&	AID (14)	\\
2459859.5674	&	2022.7640	&	{\it Vis.}->{\it V}	&	9.814	$\pm$	0.042	&	AID (13)	\\
2459860.3628	&	2022.7662	&	{\it B}	&	10.875	$\pm$	0.050	&	AID (HKEB)	\\
2459860.3632	&	2022.7662	&	{\it V}	&	9.846	$\pm$	0.050	&	AID (HKEB)	\\
2459863.3575	&	2022.7744	&	{\it Vis.}->{\it V}	&	9.837	$\pm$	0.057	&	AID (7)	\\
		\hline
	\end{tabular}

\end{table}

\subsection{$B$ and $V$ Light Curve 1866--2022}

In all, I have 6288 $B$ magnitudes 1890--2022 and 124935 $V$ magnitudes 1845--2022.  Only two of these magnitudes are limits, Herschel's non-detection in 1842 and Schmidt's deep pre-eruption limit just 2.8 hours before Birmingham's discovery of the 1866 eruption.  These observations are described in Table 1, explicitly listed in Tables 2-5, and plotted in Fig. 1.  In this Section, I will present various blow-up and superposed plots to better show what was going on.

Fig. 2 shows a close-up of the two eruption light curves, where the horizontal axis is the number of days since the time of peak light.  For this, I take the times of peak light to be JD 2402734.5 (1866.363) and JD 2431860.0 (1946.105).  The rise to peak and the peak magnitude is sharply constrained by the confident non-detection (with T CrB being fainter than 5th or 6th mag) by the great observer J. F. J. Schmidt (National Observatory of Athens) just 2.8 hours before the discovery at $V$=2.0 by Birmingham.  The eruption light curves in $V$ are identical between the 1866 and 1946 events.  The peak magnitude is close to $V$=2.0, as seen by Birmingham in 1866, and as missed by half-a-day in 1946.  The $B-V$ colour near peak is close to 0.0 mag.  The colour has increased to $+$0.5 by day 20, due to the very red contribution of the secondary star.  The times for the light curve to drop by 2, 3, and 6 magnitudes from peak are $t_2$=3.0, $t_3$=5.0, and $t_6$=12.0 days.  T CrB is one of the all-time fastest novae (c.f. Schaefer 2010; 2022c).

\begin{figure}
	\includegraphics[width=0.99\columnwidth]{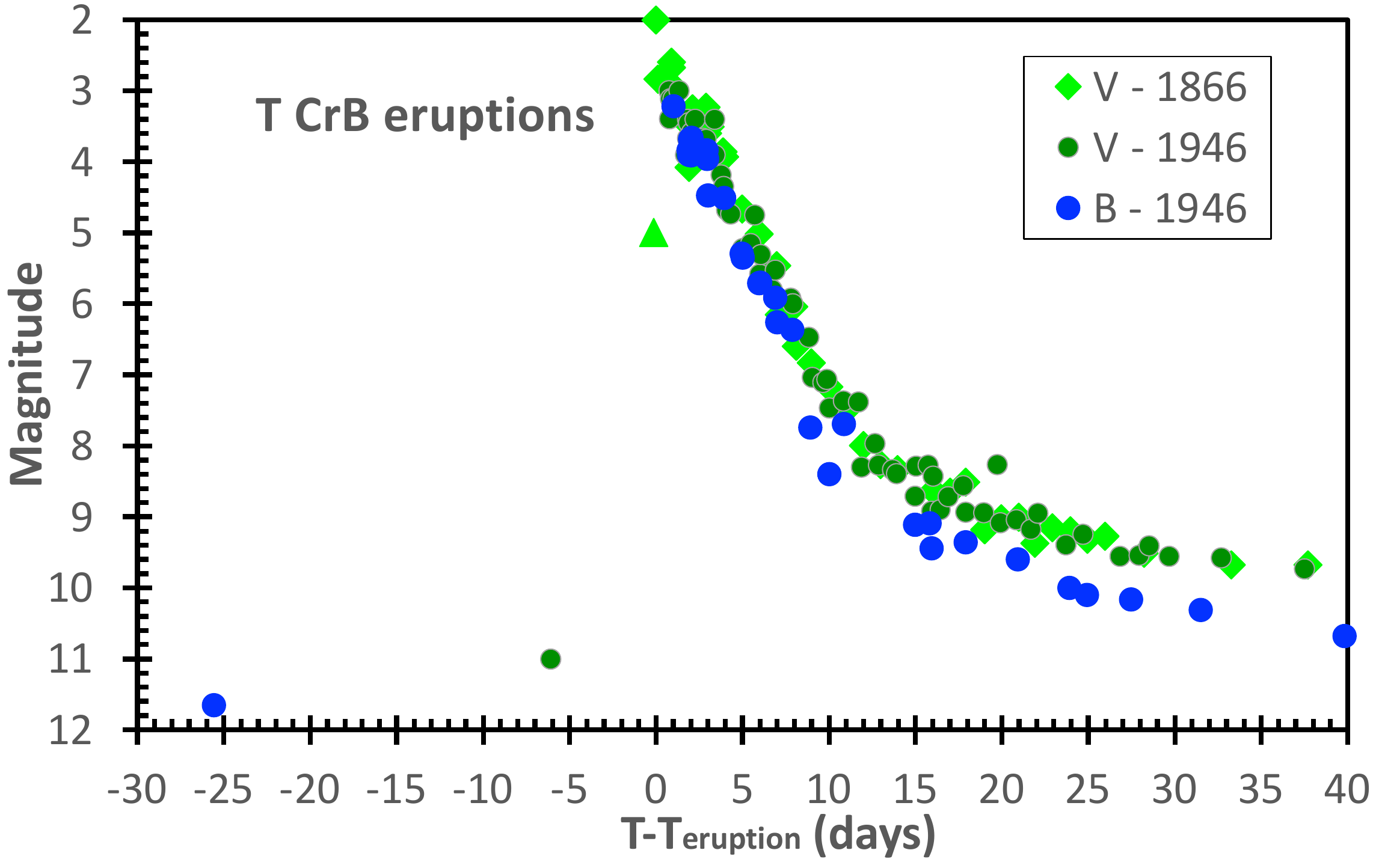}
    \caption{$B$ and $V$ eruption light curve for T CrB in 1866 and 1946.  Some binning of closely spaced magnitudes measures was used.  We see the two eruptions in 1866 and 1946 are identical in light curve shape.  The reliable null-detection of Julius Schmidt (with $V$$>$5, as shown by the light-green triangle in the figure) just 2.8 hours before the discovery observation of Birmingham at $V$=2.0 means that T CrB had a tremendously fast rise of $>$1 mag per hour.  Even with this very rapid rise, once at peak, the nova started fading at a very fast rate, falling by 3.0 mag in 5 days ($t_3$=5 days), which is one of the fastest novae on record. }  
\end{figure}

Fig. 3 shows a close-up of the time soon after the primary eruption, when the flux level goes flat at the prior level of the pre-eruption high-state.  The basic nova eruption is over and done.  Then, after an interval of over 80 days in quiescence, a secondary eruption stars.  T CrB experiences a second eruption, this with a fast rise, dominated by blue light.  The secondary event has a total duration of near 100 days, while its peak is roughly 160 days after the primary peak.   Importantly, the timing and light curve of the secondary eruption is identical between the 1866 and 1946 eruptions. This secondary eruption is unique amongst novae (and cataclysmic variables, CVs, in general), even while such events would have been detected in the case of $\sim$200 other novae, including $\sim$20 other novae with red giant companions (Strope, Schaefer, \& Henden 2010; Schaefer 2022b).

\begin{figure}
	\includegraphics[width=0.99\columnwidth]{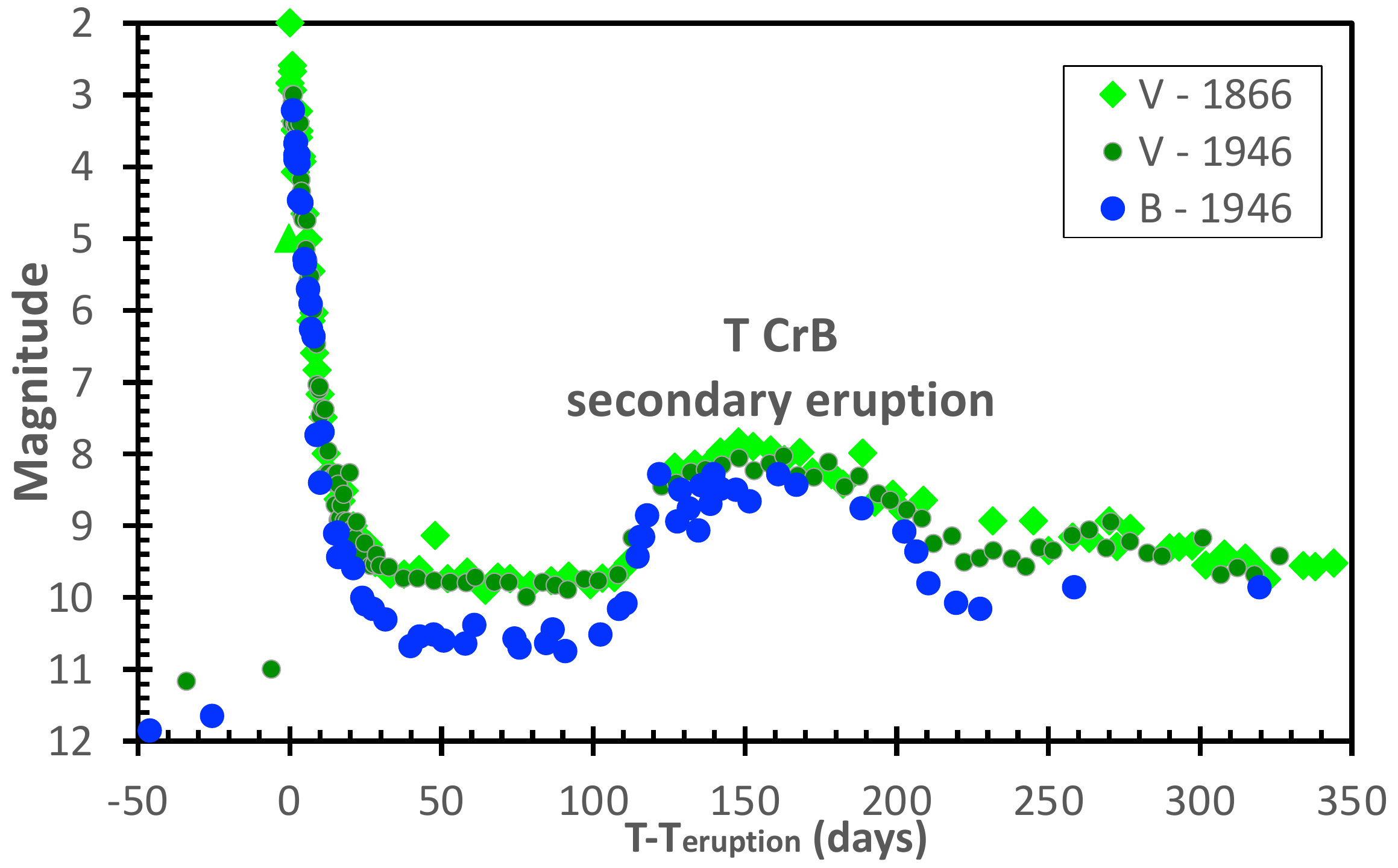}
    \caption{Secondary eruption light curve for T CrB in 1866 and 1946.  We see that the timing and structure of the secondary eruptions are identical from the $V$ band light curves in 1866 (light-green diamonds) and in 1946 (dark-green circles).  The primary eruption has entirely ended by day 30 (making T CrB by far the fastest known nova by this measure), and the light curve has returned to the prior quiescent level for 80 days.  So it is surprising that a new eruption gets somehow initiated.  The light curve shape is not that of the usual nova shape, due to the flat-top.  }  
\end{figure}

Fig. 4 shows the pre-eruption dip leading up to the 1946 eruption.  This is what L. Peltier discovered, and made him think that a second eruption was imminent (Peltier 1945).  When we see T CrB start to fade towards a dip sometime in the upcoming years, we will have advance notice for the date of the eruption.  When the dip becomes first noticeable, we will get roughly 1 year advance warning and can make a prediction of the date accurate a month or two.  In the $B$ band, the pre-eruption dip appears as a steady drop all the way to the day of the eruption.  In the $V$ band, the last year of the pre-eruption dip appears to have T CrB {\it brightening} steadily to the eruption from a minimum roughly one year in advance.

\begin{figure}
	\includegraphics[width=0.99\columnwidth]{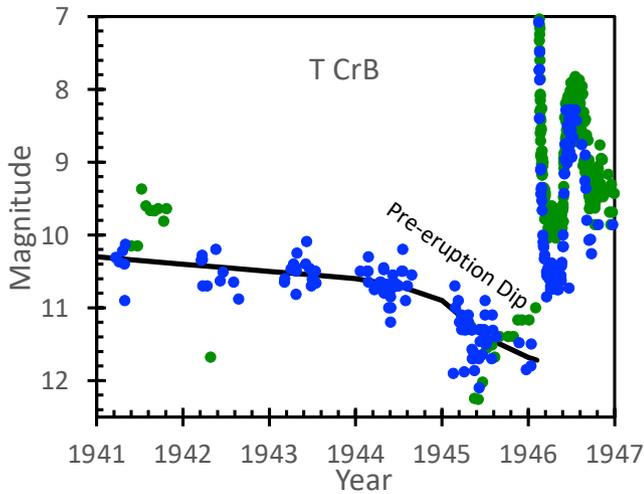}
    \caption{Pre-eruption dip in the year before the 1946 eruption.  This light curve shows all of the un-binned $B$ magnitudes (blue circles) and $V$ magnitudes (green circles) from 1941--1947.  The primary eruption runs off the top of the plot on the right side, being followed by the secondary eruption on the right edge of the plot.  In 1941, the nova had already risen from its long-term quiescent level of $B$=11.7 to its pre-eruption high-state, which has a slow fade from 10.3 in 1941 to 10.7 in 1944.  Around the start fo 1945, the nova faded down to $B$=11.7 just before the start of the 1946 eruption.  This pre-eruption dip was discovered by the great observer L. Peltier, and this phenomenon will provide a one-year warning of the exact date of the upcoming eruption.  I have not even heard of any speculation as to the physical mechanism for the pre-eruption dip.}  
\end{figure}

Away from the eruptions, the light curve of T CrB displays prominent flickering, chaotic long-term variations, and an exactly periodic ellipsoidal modulation.  The ellipsoidal effects appear as a good sinewave with exactly half the orbital period as determined from the radial velocity curve (i.e., $P$=227.5687/2=113.7843 days).  The ellipsoidal effects track the orbital position of the companion star, with maximum brightness occurring at elongations, when the red giant is viewed `sideways' with its large side showing.  These sinewave modulations are seen in my light curve all the way back to 1867, and will provide an easy means to track the orbital period and its variations 1867--2022.  Fig. 5 shows a 4-year sample from 2006--2010, with this being one of the best measured intervals.  The $V$-band light curve has the variations dominated by the ellipsoidal modulation, with a variable amplitude of 0.2--0.5 mag, as appropriate for the $V$-band light being dominated by the red giant.  Note, there is no apparent odd-even difference in the minima, which implies that any irradiation effects are relatively small.  The $B$-band light curve still shows the ellipsoidal sinewave at a similar level.  The modulation in $B$ is less well defined, likely because the chaotic addition of flickering and variations (from the accretion disc) are more prominent in blue light, plus there are much fewer $B$ magnitudes (as compared to the large number of $V$ magnitudes) to beat down the flicker variations.

\begin{figure*}
	\includegraphics[width=2.1\columnwidth]{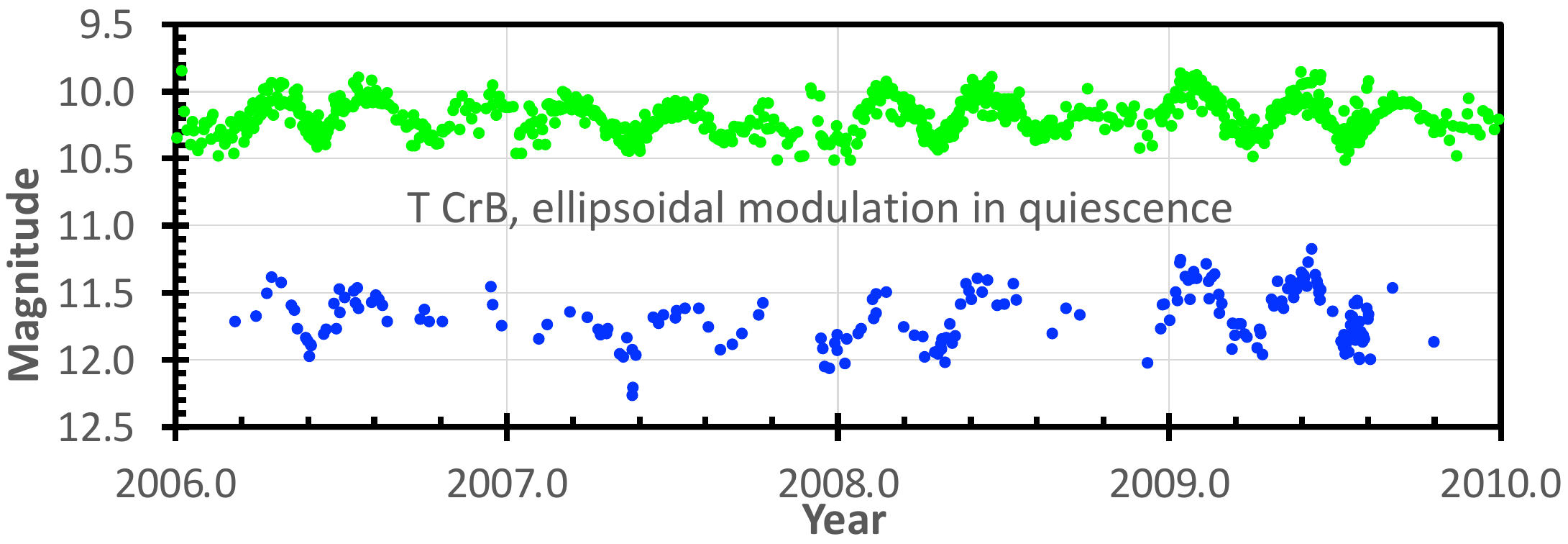}
    \caption{T CrB in normal quiescence, 2006--2010, with ellipsoidal modulation.  The red giant star is necessarily somewhat elongated due to the usual Roche lobe gravity effects from the nearby white dwarf, so as the star goes round in its orbit, it alternatively presents its broad side (and appears brightest) at orbital elongations, and its small side (and appears faintest) at conjunctions.  This ellipsoidal effect will make the light curve have a sinusoidal modulation at half the orbital period.  The resultant sinewave is easily seen in the $V$-band light curve (green dots), while the same modulations are visible in the $B$-band light curve (blue dots).  This sinewave defines the position of the companion star in its orbit, and so becomes a measure of the orbital period and its variations.}  
\end{figure*}

T CrB shows long-term variations in quiescence, as shown in Fig. 6.  For this plot, the light curves have been binned at half the orbital period (113.7843 days), so that the ellipsoidal modulations are always averaged to zero.  This also serves to minimize the variations due to flickering and short-term variations.  The changes in the level between eruptions is presumably dominated by the changes in the accretion rate.  We see that T CrB has two levels away from eruptions, which schematically are like a quantization into either a low-state or a high-state, with fairly sharp transitions.  The ordinary low level is from roughly 1875--1935 (60 years), 1955--2015 (60 years), plus at least 1855--1856.  The transitions to the high-state take roughly two years.  T CrB shows a distinct high-state, dominated by blue light, from at least 1866--1875, 1936--1954, and 2015--present.

\begin{figure*}
	\includegraphics[width=2.01\columnwidth]{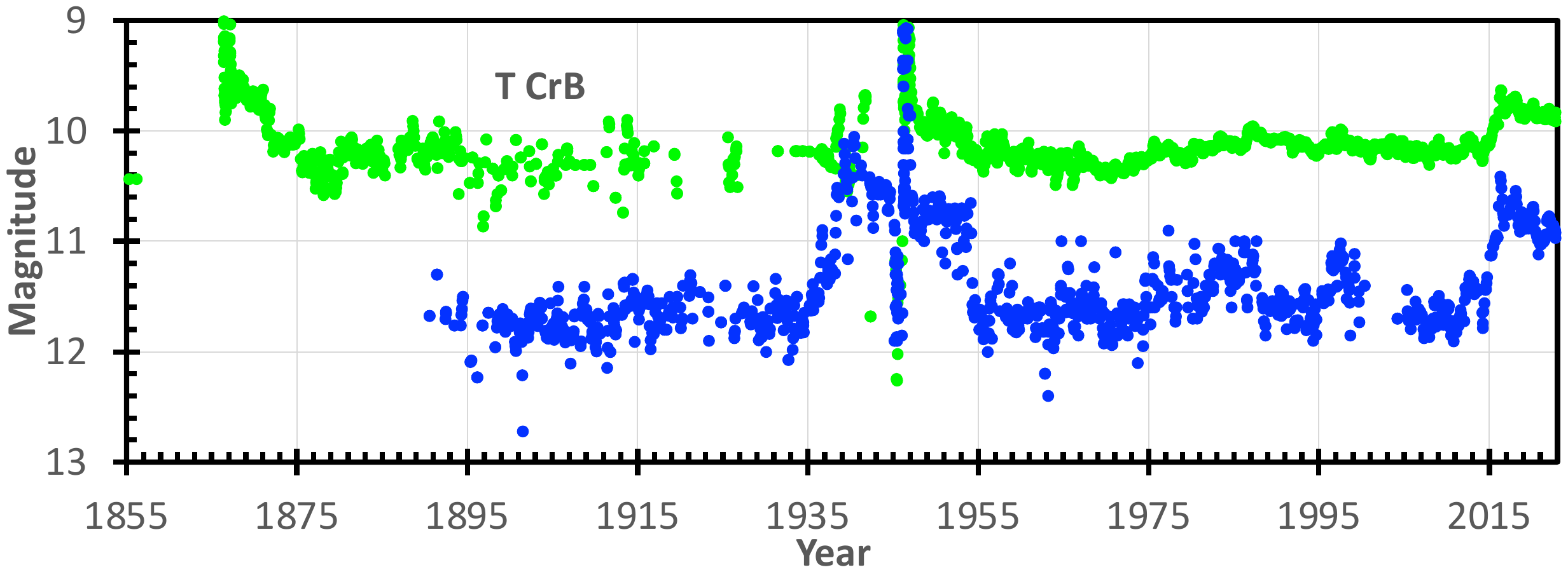}
    \caption{T CrB in normal quiescence, 1855--2022.  The $B$ magnitudes (blue circles) and the $V$ magnitudes (green circles) have been binned over a time of 113.7843 days so as to average out ellipsoidal modulations and short-term variations like flickering.  (Around the times of eruption, smaller bins sizes were used, as appropriate to show the underlying variations.)  The point of this figure is to show the high- and low-states of T CrB between eruption.  The structure of the high-state is complex, with a pre-eruption dip, the primary classical nova eruption, and the unique secondary eruption interspersed in the middle of the nearly-two-decade high-states.  Importantly, the high-state after the 1866 eruption appears identical to that after the 1946 eruption, and the high-state starting in 2015 appears similar to that starting in 1936.  With the further result that the two primary eruption light curves and the two secondary eruption light curves being identical, we have a strong case that T CrB is closely repeating itself in detail. }  
\end{figure*}

\subsection{{\it TESS} Light Curves}

{\it Transiting Exoplanet Survey Satellite} ({\it TESS}) is a mission designed to provide awesome light curves with 20--1800 second time resolution nearly continuously for many $\sim$26 day intervals for most stars in the sky down to 19th mag and fainter (Ricker et al. 2015).  {\it TESS} observed T CrB during pairs of orbits labelled as Sectors 24, 25, and 51 (see Table 1).  The time resolution was 120-s during the first two Sectors.  During Sector 51, T CrB data were returned with a time resolution of 20-s.  I have used the mission standard production of the light curves, labelled as SPOC, with these being publicly available from MAST.  The fluxes were derived with the standard `simple aperture photometry'.  T CrB is a bright star for {\it TESS}, so the flux levels (in units of electrons per second) are high and the fractional Poisson uncertainties are low.  For Sector 51, the average flux level is near 123,000 with an average Poisson level of 89.  The {\it TESS} detectors are CCDs with no filters, so the spectral sensitivity runs from 6000--10000~\AA.

Fig. 7 shows the 120-s resolution light curve for the 53.5 day interval in 2020 of Sectors 24 and 25.  We see variations on all time-scales.  The variations are at the 4 per cent level, corresponding to an amplitude of 0.04 mag.  The flickering amplitude in red light is substantially smaller than for the $V$ or $B$ bands, as expected.

\begin{figure*}
	\includegraphics[width=2.1\columnwidth]{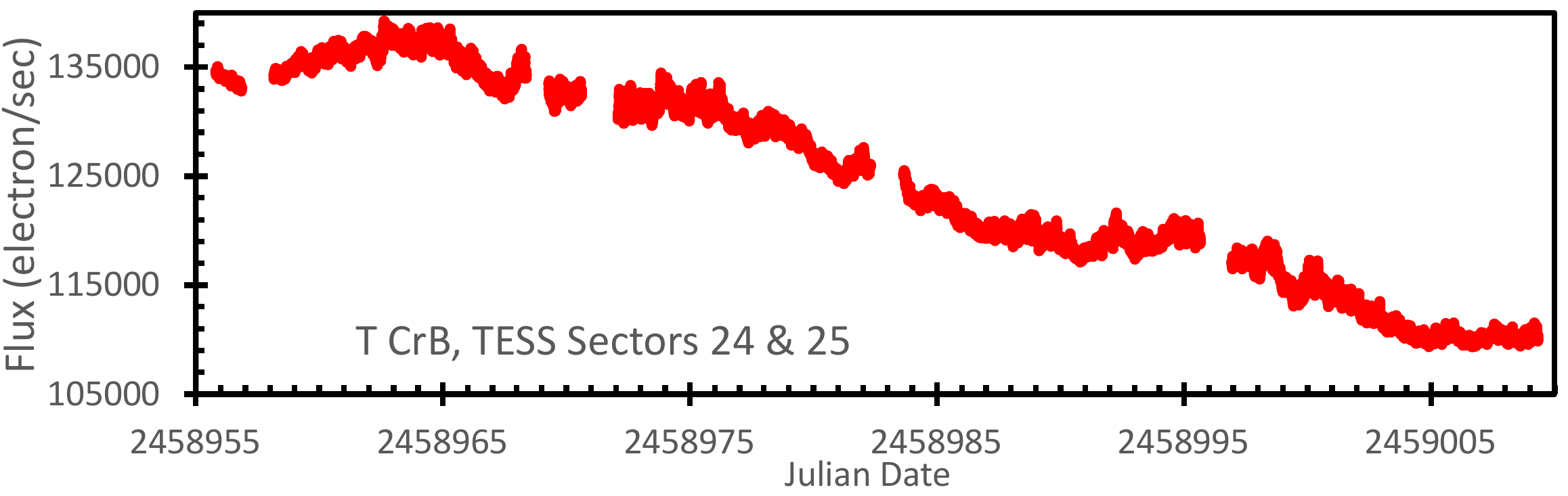}
    \caption{{\it TESS} light curve with 120-second resolution.  During Sectors 24 and 25, {\it TESS} returned 33235 fluxes over a 53.5 day interval.  The red colour of the dots is to remind us that the {\it TESS} fluxes are red sensitive, covering roughly 6000-10000~\AA.  The small gaps are caused by spacecraft operations once each perigee for data downlinks.  The Poisson error for these fluxes is $\pm$43.  With this, all of the variations in the plot are significant and intrinsic to T CrB.  We see flickering on the fastest time-scales, with a continuum of variations up to at least time-scales of several days.}  
\end{figure*}

Fig. 8 shows a one-day close-up for the Sector 51 light curve with 20-s time resolution.  Further, two insets show expanded intervals, each 0.1 days in duration.  The Poisson error bars are $\pm$86 electrons per second.  In the main figure, the Poisson errors are close to the size of each dot, so all the point-to-point variability is significant and intrinsic to T CrB.  In the insets, where each minor tick mark on the vertical axis is 200, the dots are again the same size as the Poisson errors, so the variability on the one-minute time-scale is real.

\begin{figure*}
	\includegraphics[width=2.1\columnwidth]{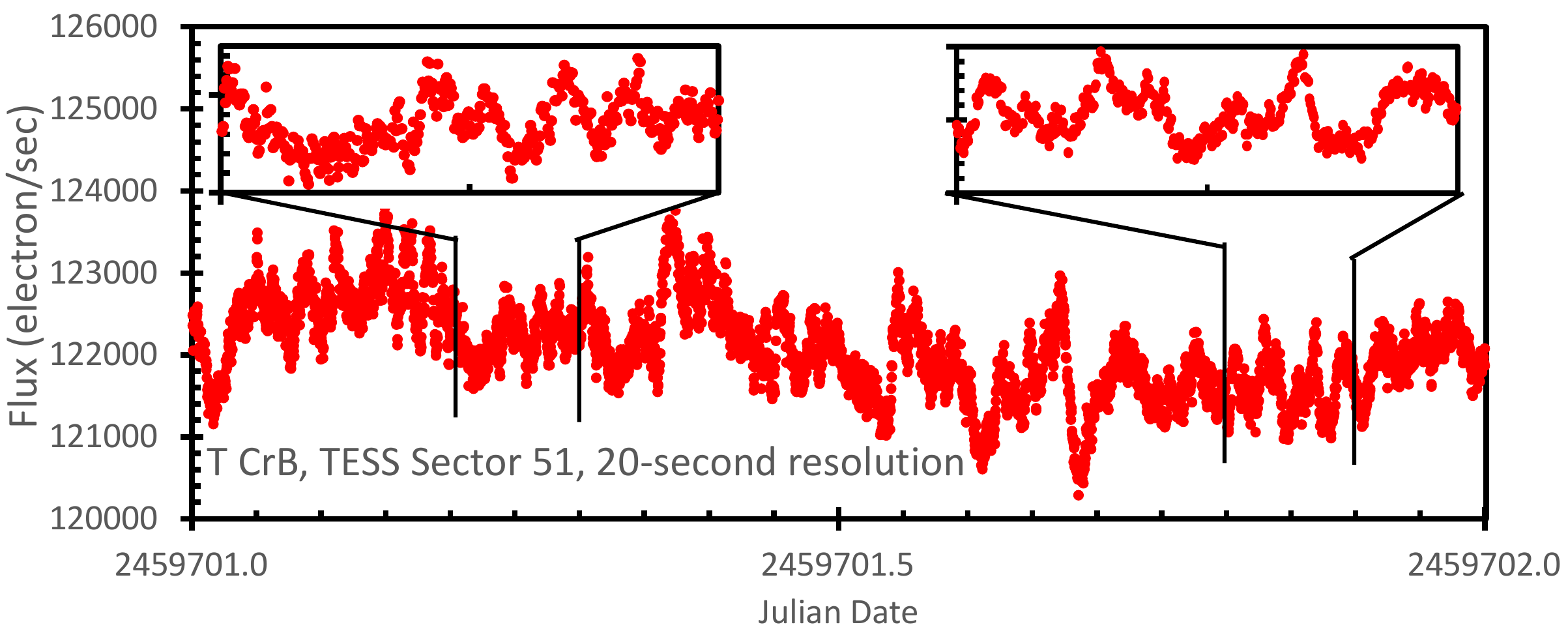}
    \caption{{\it TESS} light curve close-ups with 20-second resolution.  During Sector 51, {\it TESS} returned 20-second time resolution.  A representative sample 1.0 day interval is shown here.  The two insets show yet-finer close-ups of two 0.1-day intervals.  The Poisson error bars are 86, just somewhat larger the the plotted points.  This means that much of the variability on the one-minute and faster time-scale is significant and intrinsic to T CrB.}  
\end{figure*}

\section{THE UPCOMING ERUPTION IN 2025.5$\pm$1.3}

Many workers have recognized that the rise to a high-state, starting in 2015, is the harbinger of an eruption sometime close to 80 years after the prior eruption.  Now, with a full and definitive light curve for the years 1930--1946, plus the latest AAVSO light curve up to October 2022, we can make the best prediction as to the date of the eruption.  To this end, in Fig. 9, I have plotted a close-up of the $B$ and $V$ light curves from Fig. 1.  Further, I have used the 1930--1955 $B$ light curve to create a template of the behaviour of the eruption and high-state.  The idea is to slide the template left-right in Fig. 9, to obtain the best match in the pre-eruption high-state, then read off the year of the predicted eruption.  This presumes that the pre-eruption high-state before the upcoming eruption is the same as before the 1946 eruption.  We have seen in detail that the primary eruption light curves, the secondary eruption light curves, and the post-eruption high-state light curves are identical to within the measurement uncertainties (see Figs 2, 3, and 6), so it is reasonable to presume that the pre-eruption high-states will also be identical from eruption-to-eruption.  However, a comparison of the template versus the high-state from 2015--2022 shows (see Fig. 9) that the two are somewhat different in amplitude.  With one relatively small difference, we have the possibility that the upcoming eruption will have some small difference in timing.  The duration from 2015 to the upcoming eruption is unlikely to be greatly longer than displayed in the template, because we have the precedent from Argelander in 1856 that the duration of the pre-eruption high-state was less than ten years.

The best positioning of the template, for sliding it to earlier and later dates for the eruption year in Fig. 9, will be most constrained by the times of fastest variation.  The initial rise to the high-state is clearly defined around 1936 and around 2015.  Sliding the template right and left, I find that the initial rise is best matched for an eruption date of 2023.8, with the rise to the high-state is certainly mismatched for dates before 2023.4 and after 2024.2.  The other time when the pre-eruption high-state has a fast change is when the brightness suddenly starts fading to the pre-eruption dip.  This has not started as of February 2023.  To slide the {\it c}.1946 $B$ template such that the dip starts being apparent {\it after} 2023.2, the eruption date must be after 2024.2.

These constraints have assumed that the upcoming eruption has the same timing as the 1946-template.  But the current high-state is already different from the template by being half a magnitude fainter in $B$, so the eruption date could be longer than that seen in the pre-1946 template.  For example, maybe the timing of the eruption depends on the mass of material accreted during the pre-eruption high-state, so then the relatively small accretion over the last decade will make for an eruption date after 2023.8.  We do not know the physics of the situation, so the delay can be some unknown number of years later.  Given Argelander's measure and the close similarity of the light curves within years of the eruptions, the delay after 2023.8 cannot be longer than something like two or three years.  Given these various imperfect constraints, a final prediction would be from 2024.2--2026.8, which can be expressed as 2025.5$\pm$1.3.

After the upcoming start of the pre-eruption dip, this prediction can be substantially improved in accuracy.  On the assumption that the pre-eruption dip is similar to that of 1945--1946, the eruption will be within a few months of 1.0 years after the start of the pre-eruption dip.  On the same assumption, the eruption will be roughly three months after T CrB reaches the bottom of the dip, with an uncertainty of perhaps a month.  So, hopefully, the advance notice for the T CrB eruption might get the accuracy down to a few weeks of time.

\begin{figure}
	\includegraphics[width=1.01\columnwidth]{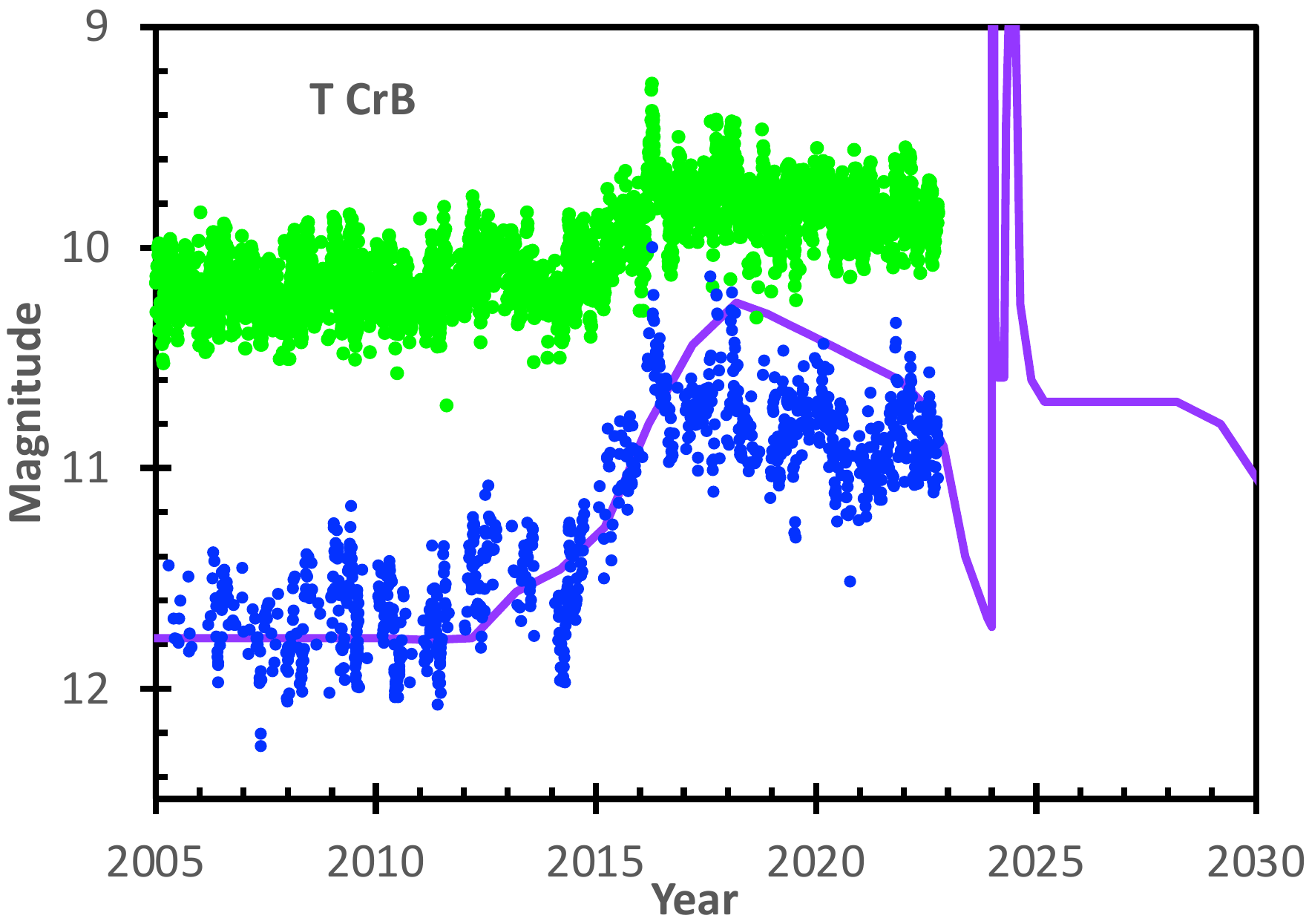}
    \caption{The upcoming eruption of T CrB is predicted for the year 2025.5$\pm$1.5.  The light curve is a close-up of that in Fig. 1, depicting the $V$ magnitudes in green and the $B$ magnitudes in blue, with data up to 2022.8.  The $B$-band template from years surrounding the 1946 eruption is shown as a purple curve, although plotted here with a shift in years so that the primary eruption peak is in the year 2024.0.  We already know that the primary eruption, the secondary eruption, and the post-eruption high-state light curves are all identical is shape and relative timing from the 1866 eruption to the 1946 eruption, so it is reasonable to expect that the shape and timing of the pre-eruption high-state are also constant from eruption-to-eruption.  Despite this expectation, this figure shows that the current high-state has a smaller amplitude than that before the 1946 eruption, and this modest difference in shape allows for possible modest differences in timing.  The figure shows the template shifted so that the rises around 1936 and 2015 overlap, which implies an eruption date of 2023.8.  But the pre-eruption dip has not started, as of February 2023, so the eruption should be after 2024.2.  That the current high-state is half a magnitude lower than the template suggests the possibility that the duration of the current pre-eruption high-state might be somewhat longer than in the template, perhaps by 2--3 years.  With this, the eruption year would be between 2024.2 and 2026.8, or 2025.5$\pm$1.3.}
\end{figure}

\section{FOLDED LIGHT CURVES}

The orbital period and the phasing of the conjunctions are known with high accuracy from the 1946--1999 radial velocity curve compiled and fitted by Fekel et al. (2000), with a period of 227.5687$\pm$0.0099 and  maximum velocity at JD 2447918.62$\pm$0.27.  The measured lines are from the atmosphere of the red giant (with no confusing lines from the accretion disc), so the radial velocity curve is a faithful measure of the actual geometric position of the star in its orbit.  The fitted orbital eccentricity is zero, with an uncertainty likely smaller than 0.012 (c.f. Kenyon \& Garcia 1986).  I have used this ephemeris to phase up the $B$ and $V$ light curves.  The phase 0.00 is for maximum velocity when the companion star is at an elongation in its orbit, on the side moving away from Earth.    At phase 0.00, the broad-side of the red giant star will be directly towards Earth, so this is the phase of maximum for the ellipsoidal variations.  At phase 0.25, the companion star is at conjunction, on the far side of the white dwarf, where we have the fullest view of the irradiated hemisphere.  At phase 0.50, the companion star is at the other elongation, where the red giant is moving towards Earth, with the Roche geometry presenting the largest cross sectional view of the companion, so this must be the phase of maximum ellipsoidal effect.  At phase 0.75, the red giant is at inferior conjunction, with the cross sectional view of the companion being minimal, and the irradiated hemisphere largely invisible.

Fig. 10 shows the visual light curve binned into phases, with the top panel for the years 1955--2015 of the low-state in quiescence, and with the bottom panel for the years 2016--2022 of the ongoing high-state.  Both panels show a folded light curve that is nearly sinusoidal with a full-amplitude of 0.17 mag (as depicted by the black sinewave).  In the low-state, the minimum at phase 0.25 (with the fullest view of the irradiated hemisphere) is not significantly brighter than the minimum at phase 0.75 (with the irradiated hemisphere largely hidden), which says that the irradiation effects are negligibly small in the $V$ band in the low-state.  The maxima occur at phases 0.11 and 0.50.  For a closely circular orbit, the maxima and minima of the ellipsoidal effects must be exactly 180$\degr$ apart.  With the T CrB maxima deviating from the ellipsoidal requirement by 36$\degr$, there must be some other effect acting to make the maximal light at an orbital phase 36$\degr$ after the elongation at phase 0.75.  Perhaps the hotspot (where the accretion stream hits the outer edge of the accretion disc) has a beaming pattern that is brightest around phase 0.20 such that the added effects of the hotspot and the ellipsoidal shape make for a maximum light at phase 0.10.  However, for this idea, the two maxima are at the same brightness, so an additional light source phased with a maximum at phase 0.50 would have to be closely equal to the light from the hotspot.  Further, my understanding is that it will be difficult for the hotspot to produce a beaming pattern that peaks at phase 0.20 or so.  With these difficulties and required coincidences and yet more required light sources, a model involving the hotspot seems poor.  In all, I have no useful idea as to the reason why the maxima are at phases 0.10 and 0.50 in the folded light curve.

The bottom panel of Fig. 10 shows the folded visual light curve for the high-state.  The high-state light curve shows maxima at phases 0.11 and 0.50.  The added light from the high-state raised the average brightness from $\langle V \rangle$=10.16 to $\langle V \rangle$=9.79.  The amplitude of the high-state light curve is close to that of the low state, so the extra light is not strongly modulated by the orbit.  However, the individual maxima and minima are somewhat unequal in the high-state.  The apparent deviations from the average amplitude for individual extrema is typical of that seen in the low state, where folded light curve for 6-year intervals always vary up and down somewhat due to the well-known ordinary fluctuations of T CrB (see Section 7).  That is, the shape, amplitude, and phasing of the $V$-band light curve are the same from the low-state to the high-state.

Fig. 11 shows the phase-folded $B$ light curve for the low-state (top panel) and the high-state (bottom panel).  The black sinewave is the schematic ellipsoidal model with an amplitude of 0.42 mag in both panels.  The low-state appears similar in $B$ and $V$, although the amplitude is larger in the $B$.  The maxima and minima are at closely similar levels, so the irradiation effects are negligible in the $B$ band in the low state.  The maxima are at phases 0.04 and 0.45, which implies that some additional effect is being superposed on the ellipsoidal and irradiation effects.  The extra light in the high-state raised $\langle B \rangle$=11.63 to $\langle B \rangle$=10.81.  The extra light still has the half-period signal that shows ellipsoidal modulation, but the light curve shape has substantial deviations from a sinewave.  I note that these deviations from the sinewave are typical of those seen over other six year intervals.  That is, the ubiquitous flickering and flaring on all time-scales makes for a light curve with relatively few cycles always appears to have deviations from the average due to the happenstance of durations and timings of the fluctuations with a time-scale of a month or so.  Thus, the deviations from a sinewave in the high-state are consistent with the ordinary and expected T CrB variability.

\begin{figure}
	\includegraphics[width=1.01\columnwidth]{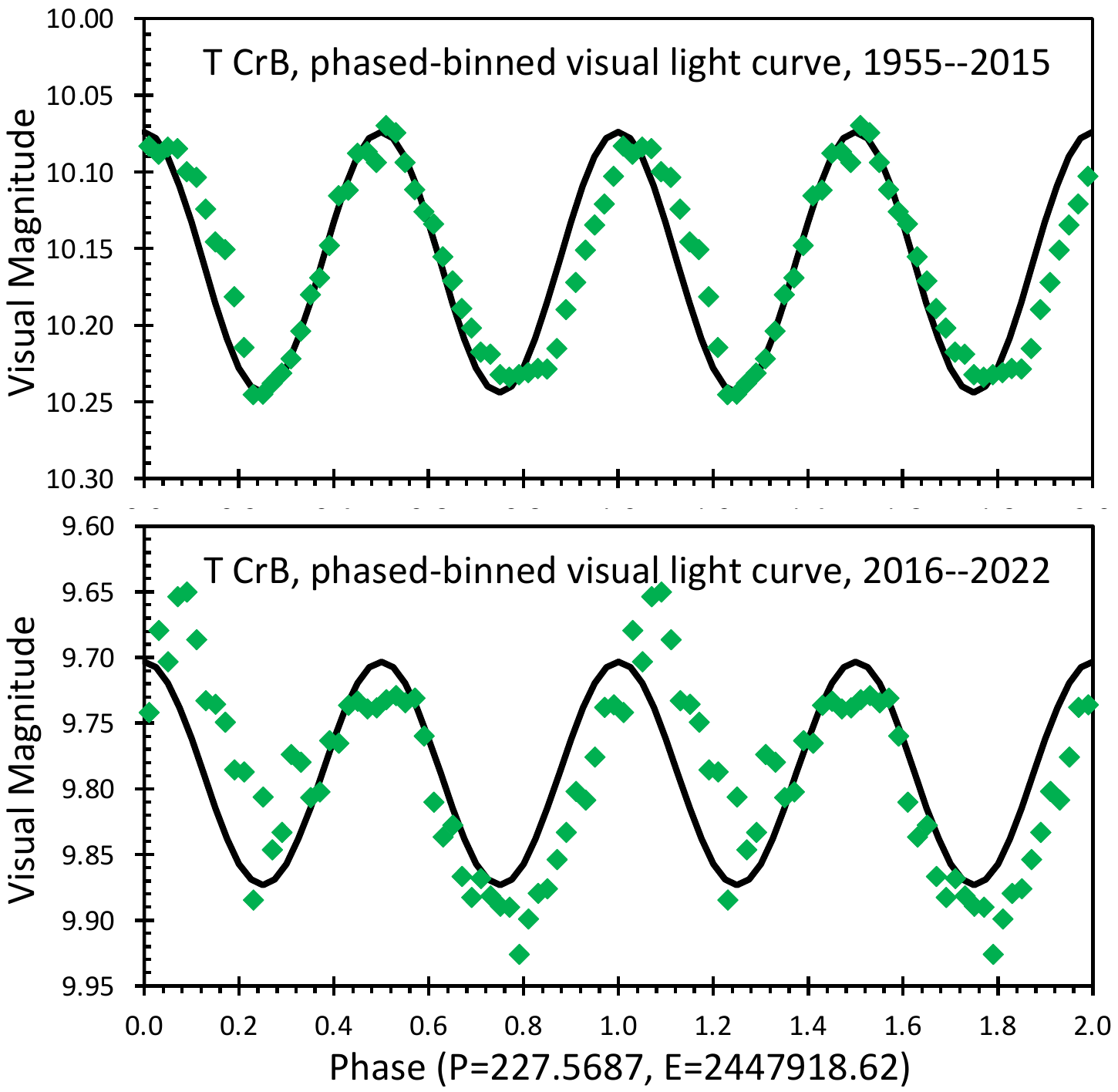}
    \caption{Phase-folded visual light curves for the low-state (top panel) and the high-state (bottom panel).  The black sinewave is the schematic effect of ellipsoidal modulations, where the maxima must be at phases 0.00 (or 1.00 in this doubled plot) and 0.50 (or 1.50).  Mysteriously, both panels show maxima near phases 0.11 and 0.50, implying some light source that is not perfectly symmetrical with orbital phase.  The top panel has averaged over 60 years (near 96 orbits), which smooths out the aperiodic fluctuations so prominent for T CrB.  The bottom panel has averaged over only six years, so small variations are seen that arise from the ordinary sampling on a source with substantial and chaotic long-term variations.}  
\end{figure}

\begin{figure}
	\includegraphics[width=1.01\columnwidth]{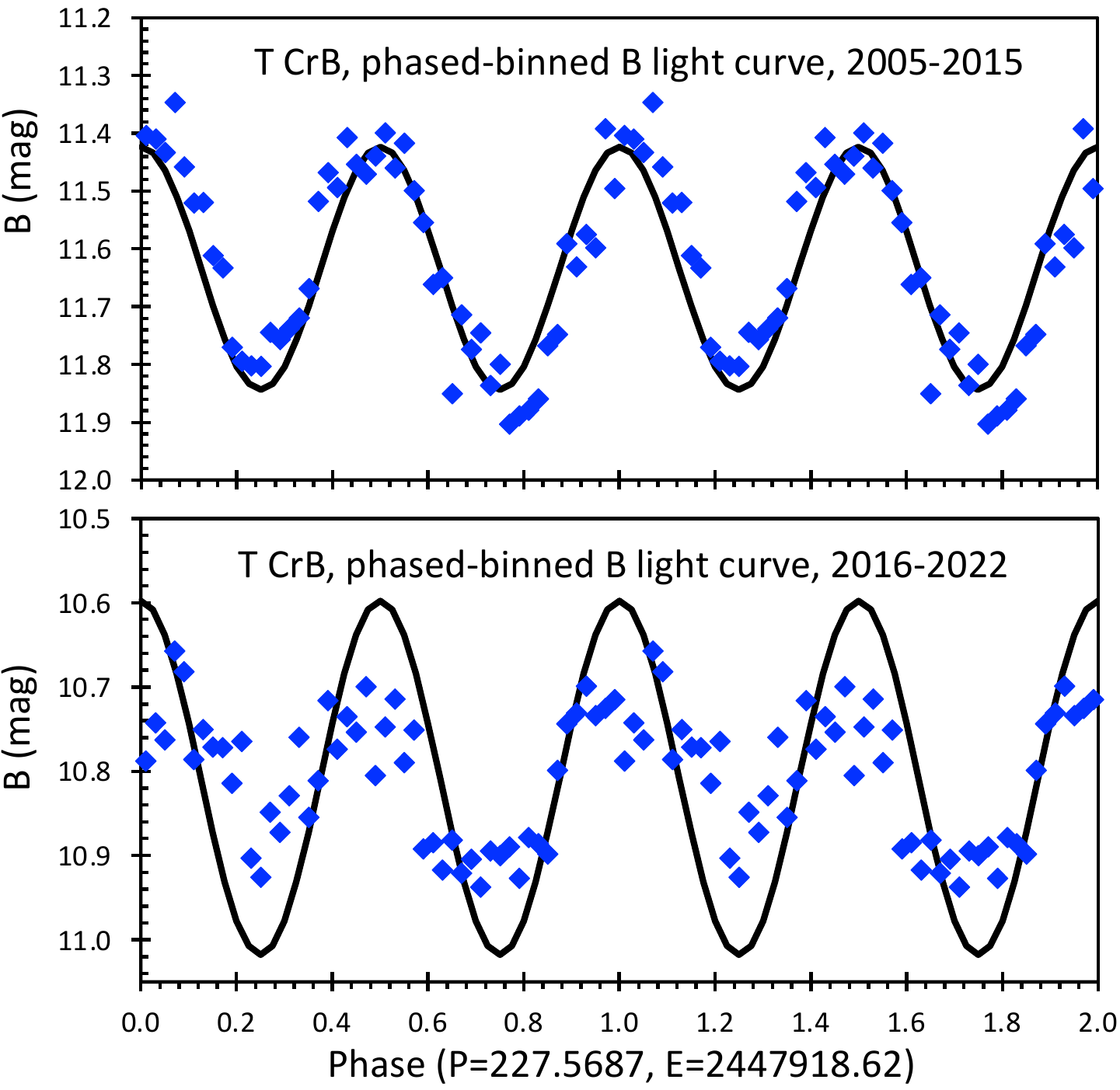}
    \caption{Phase-folded $B$ light curves for the low-state (top panel) and the high-state (bottom panel).  The black sinewaves in both panels show the schematic effects of ellipsoidal modulations, with 0.42 mag full-amplitude.  In the low-state, we see ellipsoidal modulations, with no apparent irradiation effects, although the maxima appear at unexpected phases of 0.04 and 0.45.  In the high-state, the two minima are of unequal depths (like the effect from irradiation).  In the bottom panel, the deviations from the sinewave are consistent with the sampling over a relatively short number of years with chaotic fluctuations superposed.}  
\end{figure}

\section{THE COMPLEX CHANGES IN THE ORBITAL PERIOD}

Near half of my motivation for the long work of compiling an exhaustive T CrB light curve with correct modern photometry is so that I can measure the orbital period changes all the way back to the 1866 eruption.  The idea is that the T CrB ellipsoidal modulations closely define the orbital position of the companion star (with the maxima pointing to the elongations), so monitoring the timing of the maxima will give the true orbital period as a function of time.  T CrB is the only known CV for which this is possible, because only T CrB has a large-amplitude ellipsoidal modulation and is bright enough to have a very long photometric record.  So the task is to measure times of maxima throughout the $B$ and $V$ light curves from 1866--2022, place these on a traditional $O-C$ diagram, and calculate the long-term changes in the orbital period of T CrB.

\subsection{$O-C$ Curve for 1866--2022 and Broken-Parabola Fits}

For this task, I have extracted from my primary light curve (see Tables 2--5) all the magnitudes in each colour for time intervals with durations 2--10 years.  The time intervals were chosen so as to include enough magnitudes to keep the error bars usefully small, and to avoid crossing state changes.  (The eruption years of 1866 and 1946 were not included.  Time intervals with scant data and uselessly-large error bars are not included.)  Each set of selected magnitudes was then fit to a sinewave with a period of half-227.5687 days with a standard chi-square analysis.  The time of peak is from the model fit with the smallest chi-square, selecting the peak close to zero phase in the radial velocity ephemeris that is closest to the average date of the input magnitudes.  The reduced chi-square values for these fits is always close to unity, which is to say that my quoted photometric error bars are reasonable, and this allows accurate estimates of the uncertainties in the times of minima as being the range over which the chi-square is within unity of its lowest value.  The result is 31 times of maximum light, with their Julian dates tabulated in Table 6.

\begin{table}
	\centering
	\caption{Times of maximum light and $O-C$ measures 1867.0--2022.8}
	\begin{tabular}{llllrr} 
		\hline
		Year range   & Band  &  JD maximum  &   $\langle$Year$\rangle$  &   $N$   &  $O-C$   \\
		\hline
1867	 --	1871	&	$V$	&	2403554.9	$\pm$	3.3	&	1868.608	&	-195	&	12.2	\\
1871	 --	1875	&	$V$	&	2404921.7	$\pm$	2.8	&	1872.350	&	-189	&	13.6	\\
1875	 --	1880	&	$V$	&	2406744.3	$\pm$	2.3	&	1877.341	&	-181	&	15.6	\\
1880	 --	1890	&	$V$	&	2409010.7	$\pm$	3.4	&	1883.546	&	-171	&	6.3	\\
1890	 --	1910	&	$V$	&	2414480.6	$\pm$	5.5	&	1898.522	&	-147	&	14.6	\\
1900	 --	1910	&	$B$	&	2416973.8	$\pm$	2.6	&	1905.348	&	-136	&	4.5	\\
1910	 --	1925	&	$V$	&	2420384.3	$\pm$	6.7	&	1914.685	&	-121	&	1.5	\\
1910	 --	1920	&	$B$	&	2420392.8	$\pm$	2.9	&	1914.709	&	-121	&	10.0	\\
1925	 --	1927	&	$V$	&	2424481.8	$\pm$	4.7	&	1925.906	&	-103	&	2.8	\\
1920	 --	1930	&	$B$	&	2424720.7	$\pm$	6.5	&	1926.560	&	-102	&	14.1	\\
1930	 --	1935	&	$B$	&	2426984.2	$\pm$	3.3	&	1932.758	&	-92	&	1.9	\\
1938	 --	1939	&	$V$	&	2429031.9	$\pm$	4.0	&	1938.364	&	-83	&	1.5	\\
1947	 --	1951	&	$V$	&	2432899.4	$\pm$	1.8	&	1948.952	&	-66	&	0.3	\\
1951	 --	1955	&	$V$	&	2434268.3	$\pm$	2.2	&	1952.700	&	-60	&	3.8	\\
1955	 --	1960	&	$B$	&	2436087.6	$\pm$	4.1	&	1957.680	&	-52	&	2.6	\\
1955	 --	1960	&	$V$	&	2436090.1	$\pm$	2.2	&	1957.687	&	-52	&	5.1	\\
1960	 --	1970	&	$B$	&	2438363.9	$\pm$	3.6	&	1963.912	&	-42	&	3.2	\\
1960	 --	1970	&	$V$	&	2438821.2	$\pm$	1.1	&	1965.164	&	-40	&	5.3	\\
1970	 --	1980	&	$B$	&	2442011.3	$\pm$	5.3	&	1973.898	&	-26	&	9.5	\\
1970	 --	1980	&	$V$	&	2442464.3	$\pm$	1.2	&	1975.138	&	-24	&	7.3	\\
1980	 --	1990	&	$B$	&	2445873.7	$\pm$	2.6	&	1984.473	&	-9	&	3.2	\\
1980	 --	1990	&	$V$	&	2446101.8	$\pm$	0.9	&	1985.097	&	-8	&	3.7	\\
1990	 --	2000	&	$V$	&	2449742.8	$\pm$	0.8	&	1995.066	&	8	&	3.6	\\
1990	 --	2000	&	$B$	&	2449742.9	$\pm$	1.3	&	1995.066	&	8	&	3.7	\\
2000	 --	2010	&	$V$	&	2453611.7	$\pm$	0.7	&	2005.658	&	25	&	3.9	\\
2000	 --	2010	&	$B$	&	2454518.8	$\pm$	1.3	&	2008.142	&	29	&	0.7	\\
2010	 --	2015	&	$V$	&	2456113.7	$\pm$	0.9	&	2012.508	&	36	&	2.6	\\
2010	 --	2015	&	$B$	&	2456114.5	$\pm$	1.0	&	2012.511	&	36	&	3.4	\\
2015	 --	2020	&	$V$	&	2457934.1	$\pm$	0.8	&	2017.492	&	44	&	2.5	\\
2015	 --	2023	&	$B$	&	2458612.3	$\pm$	1.7	&	2019.349	&	47	&	-2.0	\\
2020	 --	2023	&	$V$	&	2459297.6	$\pm$	1.3	&	2021.225	&	50	&	0.5	\\
		\hline
	\end{tabular}
\end{table}

\begin{figure}
	\includegraphics[width=1.01\columnwidth]{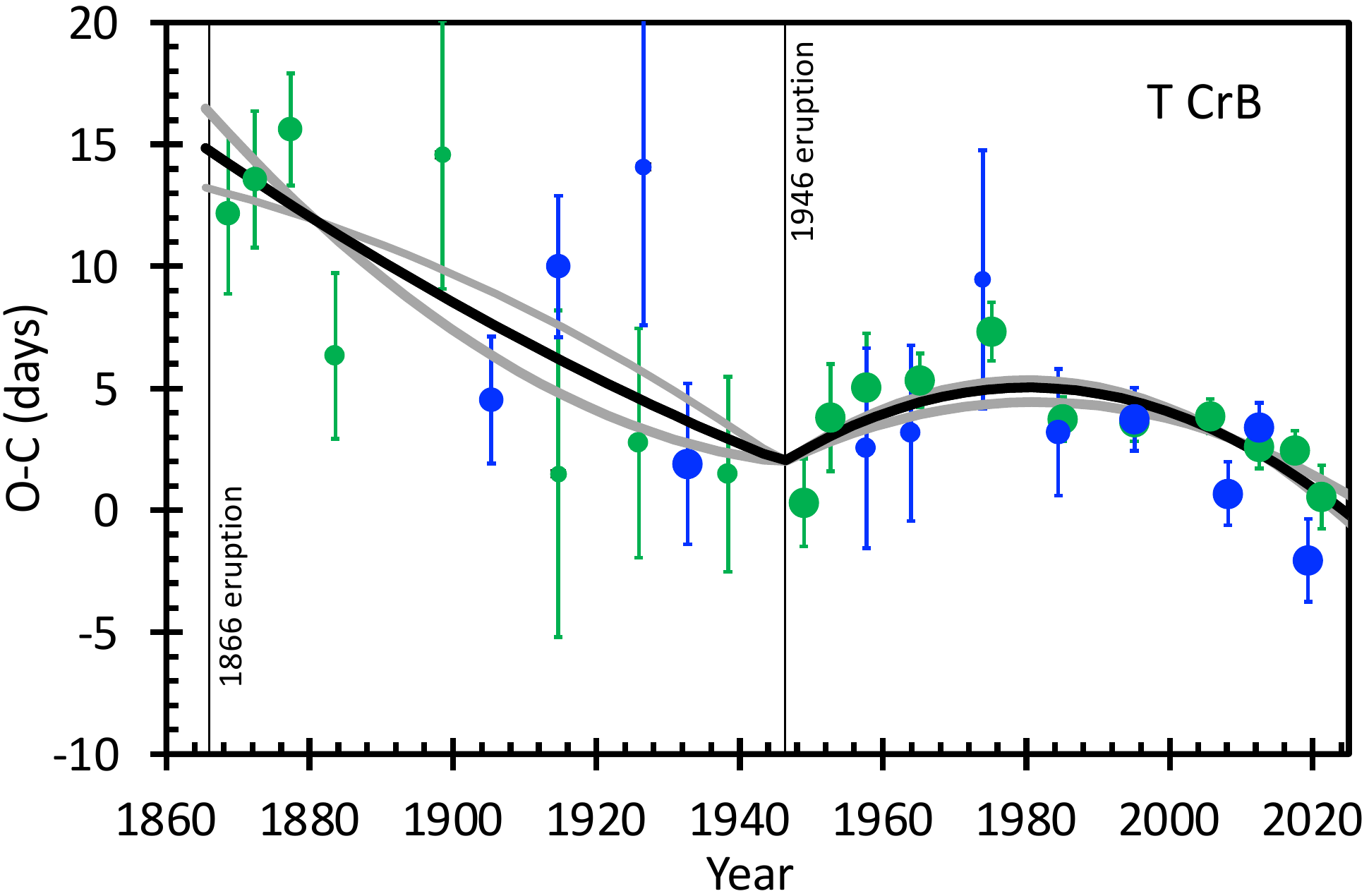}
    \caption{T CrB $O-C$ curve 1867--2022.  The $O-C$ measures from Table 6 are for the $B$ light curve (blue points) and the $V$ light curve (green points) as calculated by a chi-square fit to a sinewave, with the fiducial ephemeris for an adopted period of 227.5687 and epoch of JD 2447918.62.  The size of the points depends on the size of the error bars, with large dots for values with small error bars, with this adopted so that the visual picture is not dominated by the points with the largest uncertainty.  We see the expected $O-C$ curve as a broken parabola (thick black curve), with the break at the 1946 eruption, and the curvatures during quiescence being different between the two eruption intervals of 1866--1946 and 1946--2022.  We see good agreement between $B$ and $V$, and the calculated error bars produce a reduced chi-square satisfactorily close to unity (1.26 for 26 degrees of freedom).  The thick grey curves show the broken parabola models for extreme ranges of values that are still within the one-sigma region.  The significant period change across the 1946 eruption (seen as the kink in 1946) has $\Delta P$=0.185$\pm$0.056 days, which means that the orbital period {\it increased} due to the eruption.  The steady period change between eruption is significantly different from 1866--1946 (with a $\dot{P}$ consistent with zero) to 1946--2022 with $\dot{P}$=($-$8.9$\pm$1.6)$\times$10$^{-6}$ days-per-day.  These measured period changes are difficult to understand with published models.}  
\end{figure}

The $O-C$ curve is a listing and plot of the deviations of the observed times from some linear model times as a function of the year.  For this, some arbitrary linear ephemeris must be adopted, and I have adopted the ephemeris of Fekel et al. (2000), with $P$=227.5687 days and an epoch for maximum light of JD 2447918.62.  With this fiducial ephemeris, each of my observed times of maximum light can be assigned an integer, $N$, that counts the elapsed orbits forwards or backwards from the epoch.  With $N$, the predicted time of maximum can be calculated for each observed time of maximum, with the difference being $O-C$.  The $O-C$ values have units of days, and have the same uncertainties as the JD for the times of maxima.  These values are tabulated in Table 6, and plotted in Fig. 12.

In the $O-C$ diagram, we can only expect two types of period changes.  (Details of the construction, equations, fitting, and interpretation of the $O-C$ diagram are extensively discussed and analyzed for both observational and theoretical effects in Schaefer 2011; 2020a.)  The first type is the sudden period change that happens at the time of the eruption, where $\Delta P$=$P_{\rm after}$$-$$P_{\rm before}$, for the orbital periods just before and after the eruption.  Various competing effects are operating, including mass-loss and frictional angular momentum loss, while there also certainly must be some additional  now-unknown effect that dominates over the other known effects.  $\Delta P$ appears in the $O-C$ curve as a sharp kink, where a turn upwards means that the period suddenly increased at the time of the eruption.  The second type of period change visible in the $O-C$ curve is the slow and steady period increase (or decrease) operating throughout the entire time between eruptions.  This period change is denoted as $\dot{P}$, which is the change in $P$ over a given time period, so the quantity is in units of days-per-day, or dimensionless.  As seen in many other novae, $\dot{P}$ is apparently constant throughout the inter-eruption interval, and this agrees with the expectation that the condition of the quiescence nova binary is stable between eruptions so the period-change mechanisms should be nearly constant.  A constant $\dot{P}$ appears in the $O-C$ curve as a simple parabola extending between the times of eruptions, with a concave-down parabola showing that the period is steadily {\it decreasing}.  Surprisingly, the well-observed case of RN U Sco proves that the constant $\dot{P}$ between eruptions can change by an order-of-magnitude across the time of an eruption, so the T CrB $\dot{P}$ could easily be different for the 1866--1946 and 1946--present intervals.  In all, the T CrB 1866--2022 $O-C$ curve must appear close to a broken-parabola, with the break in the year 1946 and the parabolic curvature from before-1946 need not be the same curvature as after-1946.

I have fitted the T CrB $O-C$ curve to a broken parabola.  This best-fitting model is shown as the black curve in Fig. 12.  The reduced chi-square is close to unity, so the quoted error bars are good.  My fitted $P_{\rm before}$ is 227.495$\pm$0.053 days, while $P_{\rm after}$ is 227.680$\pm$0.019 days.  This makes $\Delta P$=0.185$\pm$0.056 days.  This sudden period change across the 1946 eruption is seen in Fig. 12 as the sharp upward kink.  Importantly, the period is getting {\it longer} across the eruption.  After 1946, the $O-C$ shows substantial curvature with a concave-down parabola, for which the best-fitting $\dot{P}$ is ($-$8.9$\pm$1.6)$\times$10$^{-6}$ days-per-day.  Before 1946, the curvature is greatly different, appearing with at most only small curvature consistent with zero, where the uncertainty is moderately large due to the relatively large error bars on the pre-1946 $O-C$ input.  The pre-1946 $\dot{P}$ is ($+$1.75$\pm$4.5)$\times$10$^{-6}$ days-per-day.

The null-hypothesis is no period change, i.e., a straight line model in Fig. 12.  The confidence in adding a parabolic term can be estimated from the change in chi-square (Lampton, Margon, \& Bowyer 1976).  For this, the chi-square increases by 17.1 over its minimum, which is to say that the null-hypothesis is rejected at $>$4-sigma confidence level.  The null hypothesis for $\Delta P$ is a value of zero, which is a single parabola model.  For this, the chi-square is 16.6 larger than the minimum, which is to say that this null hypothesis is rejected at $>$4-sigma.  The null hypothesis for $\dot{P}$ is a value of zero (making a model as a broken line), which has a chi-square that is 13.3 larger than the overall minimum, so the possibility of no-curvature is rejected at the 3.6-sigma confidence level.    For the curvature, we can look at the post-1946 $O-C$ points, with this time interval having the values with the smallest error bars and no complications from $\Delta P$ or changing $\dot{P}$.  For this interval, the fitted $\dot{P}$ is (-7.8$\pm$2.5)$\times$10$^{-6}$ in dimensionless units of days per day, while the fits with no curvature are worse by 11.2, which indicates the curvature is significant at the 3.3-sigma confidence level.  In all, the existence of the sharp period-change in 1946 is significant, and the existence of the steady period change after 1946 is significant. 

The measured $\Delta P$ and $\dot{P}$ values are greatly different from zero, so much so that theory has a difficult time explaining the sizes of both (see next Section).  In such a case, it is prudent to consider whether the basic result can be impeached in some way.  But the data are simple,  good quality (to within the stated error bars), and multiply redundant, so I see no chance to impeach the input.  The analysis is simple and standard, while the best-fitting models have reduced chi-squares of close to unity.  The existence of the period changes is measured to have a confidence level of $>$4-sigma.  So the case for large non-zero $\Delta P$ and $\dot{P}$ is convincing.  I can think of only one evidenced argument to impeach anything, and that is against just one $O-C$ value.  The $B$-band light during the high-state (see bottom panel of Fig, 11) has a different shape than during the low-state.  In this case, it is possible (but I have no other evidence for such) that the added light in the high-state is not symmetric in phase, thus introducing a systematic offset in the fitted phase of maximum light that does not reflect the position of the companion star in its orbit.  This line of speculation is applicable only to the $B$ light curve and only during the high-state, so only one $O-C$ value (the 2015--2023 $B$ value in Table 6) might be impeached.  With this one point tossed out, the best-fitting broken parabola and the best-fitting straight line are closely the same as in the previous paragraph, with the chi-square difference of 15.3, for a significance of the existence of the two types of period changes at the 3.9-sigma level.  The speculation about the symmetry of phasing of $B$-light in the high-state cannot be transferred to the $V$-light in the high-state, because the extra $V$-light in the high-state is small (see Fig. 6) and because the folded light curve is the same shape as during the low-state, to within the normal variations imposed by the flickering (see Fig. 10).  So there are no more $O-C$ curve points that can be suspected even by speculation.  In all, the existence of the large $\Delta P$ and $\dot{P}$ values is significant at the 4-sigma confidence level, and the result cannot be impeached.

\subsection{Broad Implications For These Measured Period Changes}

T CrB has a large $\Delta P$, a very large $\Delta P$.  The value for T CrB is $>$100$\times$ that of any other nova system, for which I have measures for 12 nova eruptions (e.g., Schaefer 2020b; 2022a).

A measure of this huge period change is $P/\Delta P$=1230 eruption cycles, as a schematic doubling time-scale for the orbital period.  With $\tau_{rec}$ of 80 years, the doubling time-scale is 98,000 years from $\Delta P$ alone.  This is greatly faster than all evolutionary time-scales for other known CVs.  The effective long-term period change is $\Delta P$/$\tau_{rec}$=$+$6.3$\times$10$^{-6}$,  with this being {\it opposite} the effects of $\dot{P}$ and dominating over the effects of $\dot{P}$ (with $\dot{P}$ during a single eruption cycle averaging to $-$3.6$\times$10$^{-6}$).  So the unknown mechanism that generates $\Delta P$ is dominating the evolution of T CrB.

With the abrupt increase in $P$ by 4.44 hours, the orbital semi-major axis will suddenly get larger by 0.106 R$_{\odot}$, and the companion star's Roche lobe will expand by 24,800 km.  This can be compared to the red giant's atmospheric scale height of 468,000 km.  With this, we understand that the changing Roche lobe size will not make any substantial change in the quiescent accretion rate from intereruption-to-intereruption intervals.  

This expansion of the Roche lobe by 24,800 km can also be compared to the expected expansion of a red giant star around the time when its radius is 66 $R_{\odot}$.  The exact value for the red giant expansion is uncertain, because there must certainly have been some mass transfer in the original binary, because the star that is now the companion has a mass near 0.81 M$_{\odot}$ (for a 1.35 M$_{\odot}$ white dwarf with a mass ratio of 0.60, Belczy\'{n}ski \& Miko{\l}ajewska 1998), yet must have had a somewhat larger main sequence mass so that the core can now start with the red giant expansion.  For any reasonable evolution track in the HR diagram, the red giant in T CrB will expand at a rate of near 0.5 km per year.  The evolutionary expansion of T CrB is negligibly small for time-scales of faster than a million years or so.  

The observed $\Delta P$ can be compared to values predicted for all known mechanisms (Schaefer 2020a).  The first mechanism is mass loss from the nova ejecta, which will necessarily increase $P$ by approximately $2PM_{\rm ejecta}/(M_{WD}+M_{comp})$, which is 0.000021 days for an ejecta mass of M$_{\rm ejecta}$=10$^{-7}$ M$_{\odot}$ (Selvelli et al. 1992).  This mechanism cannot account for the observed $\Delta P$ by a factor of 10$^4$$\times$.  The second mechanism is termed `frictional angular momentum loss', where the velocity of the companion star moving within the expanding nova shell is slowed down by dynamical friction with the shell's mass, hence lowering the $P$.   Detailed calculation gives the period change as $-$1.3$\times$10$^{-8}$ days.  In any case, this negligibly small effect is always {\it negative} for a decreasing period, so this second effect cannot account for the observed $\Delta P$.  The third mechanism is essentially magnetic braking of the companion star inside the expanding nova ejecta shell.  This effect will necessarily be small due to the relatively high-speed and low-mass of the ejecta, so the resultant period change is negligibly small even compared to the mass-loss effect.  In any case, this third effect is always negative for the case in hand, so it cannot account for the observed $\Delta P$.  The fourth mechanism to change $P$ across a nova eruption is to invoke asymmetric ejecta, where the ejection will produce a reaction force back on to the white dwarf.  All nova shells show substantial deviations from spherical symmetry, such that if the white dwarf in T CrB happened to have ejected an excess of mass in the backward direction of its orbital motion, then the orbital velocity will speed up and the period will suddenly increase.  For expected conditions for T CrB (say, with Selvelli's M$_{ejecta}$ and the asymmetry factor of 0.5), the calculated period change is 0.0014 days, which is greatly smaller than observed.  But if the M$_{\rm ejecta}$ is raised to near 10$^{-5}$ M$_{\odot}$, then the observed $\Delta P$ can be reproduced.  Such a high M$_{\rm ejecta}$ is not outlandish, as my estimated mass accreted between eruptions is 1.2$\times$10$^{-6}$ M$_{\odot}$, and additional mass from the surface of the white dwarf might be dredged up and ejected by the nova event.  Further, extreme asymmetries in the ejecta can increase the period change by a factor of 4$\times$.  In all, the only known mechanism with any possibility of producing the large value of $\Delta P$ is the asymmetric ejection of the nova shell, and only then if pushed to unlikely extremes.

T CrB has a large $\dot{P}$ after the 1946 eruption, a very large $\dot{P}$.  The value for T CrB is $>$200$\times$ larger than for any other known novae, for which I have measures for 14 novae (e.g., Schaefer 2020b; 2022a).  

The large $\dot{P}$ can be compared to the theoretical values predicted by various physical mechanisms.  Schaefer (2020a) summarizes all known mechanisms to produce $P$ changes between eruptions, and find that only three mechanisms are possible to make a steady $\dot{P}$ on long time-scales.  The first mechanism is that ordinary mass transfer from the companion star to the white dwarf will change the period.  The resultant $\dot{P}$ equals $3P(1+q)\dot{M}/M_{\rm comp}$, with $q$ being the mass ratio (M$_{\rm comp}$/M$_{\rm WD}$) and the accretion rate is $\dot{M}$.  For the high-state accretion, $\dot{P}$ is 5.9$\times$10$^{-8}$, which is 150$\times$ smaller than the value from the 1946--2022 interval.  In any case, this mass-transfer mechanism will always {\it increase} the period, so this cannot explain the observed $\dot{P}$.  The second $\dot{P}$ mechanism is labelled `magnetic braking', where the stellar wind from the companion is entrained to the companion's magnetic field and forced to co-rotate, so rotational angular momentum goes to the wind atoms which are then lost from the system, with tidal effects forcing synchronized rotation which then takes away angular momentum from the orbit, providing a steady negative-$\dot{P}$.  For the case of T CrB, the likely tidal synchronization of the companion star's rotation forces it to have a rotation period of 227 days, and this long rotation period means that any dynamo effect will produce only a very small magnetic field, while the slow spinning of the companion star means that any stellar wind gas will carry little angular momentum (Privitera et al. 2016).  Between these two effects, there is no real chance that magnetic braking can account for the very large $\dot{P}$ after 1946.  The third $\dot{P}$ mechanism is the gravitational radiation emitted by all binary stars, with the associated angular momentum loss slowly grinding down the orbital period.  For the case of T CrB, the orbital period is so large that the gravitational radiation effect is infinitesimal.  In all, we are left with the only known mechanisms for producing long-term steady period changes between nova eruption are all greatly too small to explain the post-1946 $\dot{P}$.

Despite having no theory explanation for the period in T CrB, we do have an empirical measure of their changes.  The $\Delta P$ dominates, with the period increasing by 4.44 hours across the 1946 eruption, so the binary is {\it separating} and the companion's Roche lobe radius is enlarging by 24,800 km each eruption.  This creates two related problems, one with the past of the T CrB system, and the other with the future of the T CrB system.  With the Roche lobe now expanding fast, we are faced with the T-CrB-progenitor problem.  That is, how can the progenitor have arrived at a situation  with a substantially smaller Roche lobe in the past, only to now be expanding?  Further, in the past, the $\dot{M}$ would have been much higher, pushing the accretion at least into the `steady hydrogen burning' state, and no such system with a red giant companion is known, despite being luminous and prominent in our Galaxy.  I have no useable answer for this T-CrB-progenitor dilemma.  The future of T CrB is apparently to separate, with the Roche lobe expanding, the accretion decreasing, and ultimately leading to a disconnected binary if continued.  However, the situation for changes in the current rates and the effects of an expanding Roche lobe are complex and unknown, so such an extrapolation by many eruption-cycles has little pretense to accuracy. 

The {\it changes} in $P$ controls the evolution of CVs in general.  For T CrB, all the known mechanisms to explain $\Delta P$ and to explain $\dot{P}$ fail by orders of magnitude.  With this, we have no useful theory for understanding the evolution of T CrB in the past or future, or even what currently powers the high-$\dot{M}$ that makes the system into an RN.

\section{THE FOURIER TRANSFORMS}

The AID and the {\it TESS} light curves have large numbers of brightness measured that are well-sampled, with these being good for searching for periodicities over a very wide range.  The ellipsoidal modulations, with a period of 113.7843 days is already well-known.  A signal at the orbital period is also possible, for example, due to irradiation effects on the companion star.  Another possible signal arises from various effects tied to the white dwarf rotation period.

The optimal way to search for strictly periodic signals is the Fourier transform.  For my period search,  I use the discrete Fourier transform program VSTAR, which also produces a fitted semi-amplitude for each trial frequency.  I have applied the Fourier transform to the unbinned AID visual light curve after 1948 (112560 magnitudes), to the {\it TESS} Sector 24 and 25 light curve of 53.5 days with 120-second resolution (33235 fluxes), and to the {\it TESS} Sector 51 light curve of 19.3 days with 20-second resolution (49272 fluxes).  The amplitude of a signal can be derived from the peak power, and the limit on any amplitude over a range of frequency can be derived from the power in the envelope of the noise in the power spectrum.

The half-orbital period produced a highly significant peak in the Fourier transform for the AID data.  This peak was at a period of 113.77 days, with a full-amplitude of 0.149 mag.  The full orbital period also appears with a peak at 227.27 days, corresponding to a full-amplitude of 0.042 mag.

I have not found any other significant non-artefact periodicities.  The transform of the AID light curve places a limit on the presence of any strict periodicity from 10--4000 days to have a full-amplitude of under 0.03 mag.  All three Fourier transforms limit the presence of coherent periodicities from 0.01--10 days to have a full-amplitude of under 0.0051 mag.  For periods from 40-seconds to 0.01 days, the transform of Sector 51 places a limit on the amplitude of any possible periodicity to be under 0.00072 mag.

\section{THE POWER SPECTRAL DENSITY}

The power spectral density (PSD) is the Fourier power (the amplitude squared) as a function of the frequency.  This is usually displayed on a log-log plot with logarithmic spacing of frequency bins.  This shows the relative strength of fast variations versus slow variations.  In CVs, the PSD appears as a noisy power-law, quantified by a power law index, $\gamma$, where the average Fourier power scales with frequency ($f$ in units of cycles per day) as $f^{-\gamma}$.  Typical values for CVs are 0.5$<$$\gamma$$<$1.5 with substantial variability (e.g., Bruch 2022).  Power-laws are always cutoff at some low- and high-frequency due to some limitation of the physical mechanism intrinsic to the star.  For CVs, the high-frequency cutoff to a steep slope is close to $\log[f]$=$+$1.9 (Scaringi et al. 2015).  Further, the observed power-laws will suffer breaks at high-frequency due to measurement noise (like from Poisson fluctuations) and will suffer from large noise at low-frequency due to the happenstance of the particular random realization of the few measures of the slowest variations.  The variations on the faster time-scales (say, faster than a few hours) is commonly termed `flickering'.  The underlying physical mechanism is still unknown, but it undoubtedly arises from fluctuations of the accretion process associated with the disc.

T CrB flickering (c.f., Figs 7--8) has been extensively measured, e.g., Walker (1977), Zamanov \& Bruch (1998), and I{\l}kiewicz et al. (2016).  Flickering on T CrB is indistinguishable from flickering on the RNe (Schaefer 2010) and on other CVs (Zamanov \& Bruch 1998), despite the large range of disc sizes.  Zamanov et al. (2004) has well-measured the T CrB PSD with 27 nights of long photometry runs in the $U$-band, concluding that the power scales as $f^{-1.46\pm0.17}$, for 1.85$<$$\log[f]$$<$3.16.  I have reported the PSD of T CrB for $-$4.3$<$$\log[f]$$<$$-$1.58 (Schaefer 2010, fig. 66) based on the 1947--2009 AAVSO visual light curve, showing a power-law of $f^{-0.9}$ with much scatter.

I have constructed a new PSD for T CrB from my complete $V$-band light curve in magnitudes for the years 1868--1945 and 1948--2022 (avoiding times near the eruptions), from the 53.5 day interval for {\it TESS} Sectors 24 and 25 with 120-s time resolution flux light curve, and from the 24.6 day interval for {\it TESS} Sector 51 with 20-second time resolution flux light curve.  The time intervals near the eruptions are excluded because the eruption light curves are not a measure of the accretion processes highlighted by the PSD.  From the $V$-band PSD, I have removed the frequencies near one-month and one-year due to the presence of the usual artefacts.  Poisson noise turns the PSD flat for $\log[f]$$>$2.33, so this is my limit.  I have added in a power-law to represent the observed $U$-band magnitude light curve PSD of Zamanov et al. (2004).  The four PSDs all have separate normalizations that are largely unknowable, but fortunately three have large overlap in frequencies, while the $V$-band PSD can be matched to the {\it TESS} PSDs at $\log[f]$ around $-$1.1.  Each PSD is normalized so that the deviations in the overlap region are minimized.

Fig. 13 shows my combined PSD covering -4.69$<$$\log[f]$$<$3.16.  This is 7.85 orders of magnitude in frequency.  The usual high-frequency break at $\log[f]$=$+$1.9 cycles per day is not seen.  It is striking that the entire PSD is close to a single power-law, with $f^{-1.22\pm0.08}$.  The existence of this PSD so close to a single power-law is consistent with the possibility that a single physical mechanism produces the variations on time-scales from 1.0 minutes to 134 years.  

\begin{figure}
	\includegraphics[width=1.01\columnwidth]{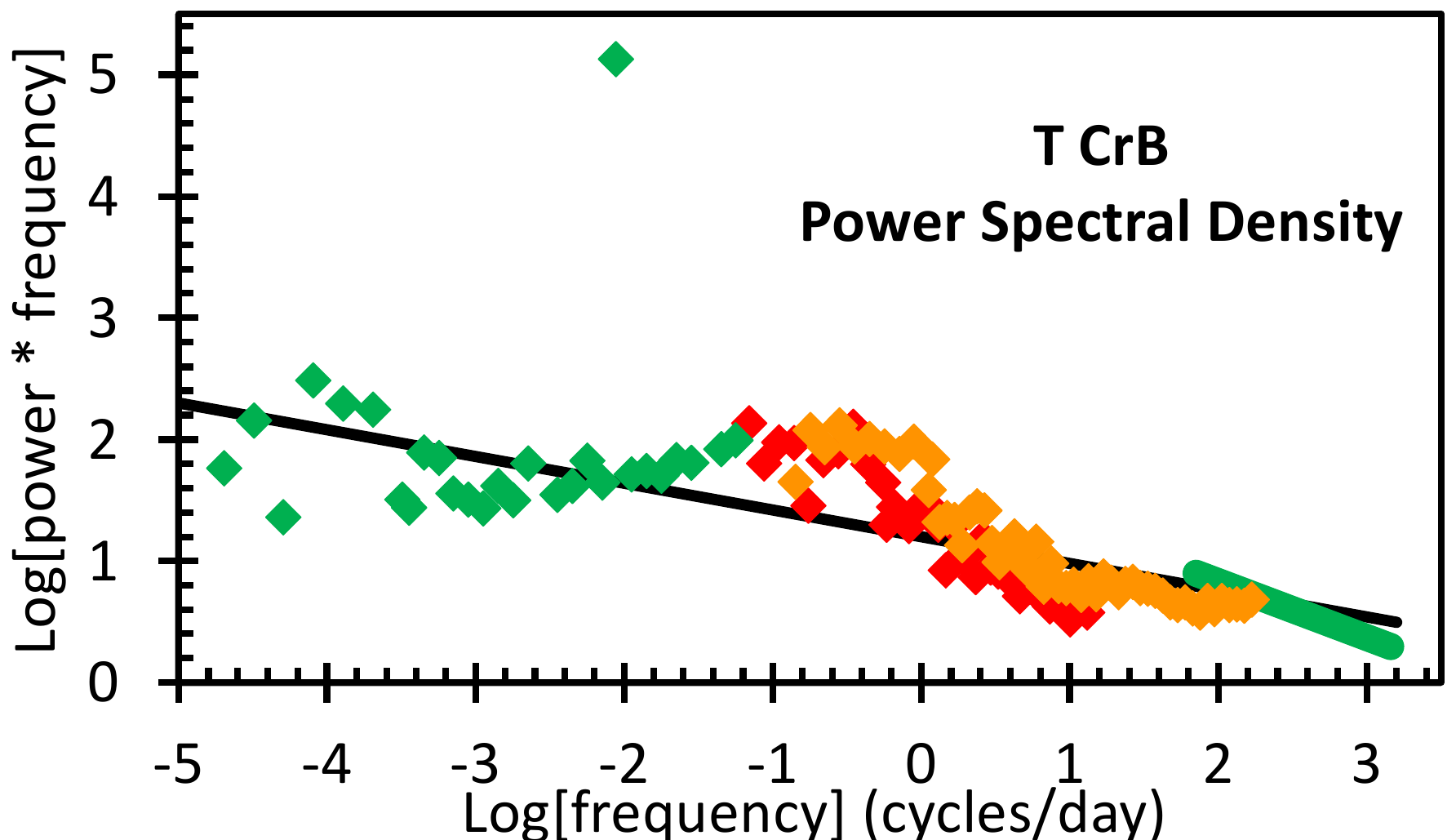}
    \caption{Power spectral density over 7.85 orders-of-magnitude in frequency.  This PSD is a composite of four normalized PSDs, derived from my $V$-band light curve 1848--1944 and 1948--2022 (green diamonds), 53.5 days of the {\it TESS} Sectors 24 and 25 (red diamonds), the 24.6 day light curve from {\it TESS} Sector 51 with 20-s time resolution (orange diamonds), and the 27-night $U$-band light curve from Zamanov et al. (green line sticking out at the lower right).  The single high point is from the ellipsoidal modulations.  The PSD is close to a single power-law with $f^{-1.22\pm0.08}$.}  
\end{figure}

\section{THE SPECTRAL ENERGY DISTRIBUTION}

The spectral energy distribution (SED) is the monochromatic spectral flux density (F$_{\nu}$ with units of Jansky), or alternatively flux ($\nu$F$_{\nu}$), as a function of the photon frequency ($\nu$ in units of Hertz), and is often displayed on a log-log plot.  The SED shows where the energy comes out, and viewers can readily see the shapes (like blackbodies and power laws) with the physical interpretations.  For T CrB, I have collected measured fluxes in Table 7 from many sources, with these being ordered by year of observation.  Most sources are from the years 1976--2012, while T CrB was in its quiescent low-state (top panel of Figure 14).  The last two source (AID and {\it Swift}) are from 2019--2022, with T CrB in its high-state (bottom panel of Fig. 14).  The X-ray data are not displayed, as the points are isolated far off the right side and far off the bottom of the plot.

The SED plot shows a good blackbody shape for the red giant companion star, plus a flattening in the ultraviolet and $U$ bands.  I have fitted a blackbody plus an $\alpha$-disc model (Frank, King, \& Raines 2002), with the fit from disc-plus-blackbody shown as a thick purple curve.  The blackbody alone is shown as a narrow grey line that merges with the total curve redward of the $B$ band (because the red giant dominates greatly over disc light).  The standard $\alpha$-disc model gives the flux across frequencies (Frank et al. 2002) where I have adopted a distance of 914 pc (Schaefer 2022c), $E(B-V)$=0.065 from {\it Galex}, a white dwarf mass of 1.35 M$_{\odot}$ (Shara et al. 2018; Hachisu \& Kato 2019), a mass ratio of 0.60 (Belczy\'{n}ski \& Miko{\l}ajewska 1998), a disc size of 22 per cent of the white dwarf Roche lobe size (Eq. 4.20 Frank et al. 2002), an orbital period of 227 days, and an orbital inclination of 60$\degr$  (Belczy\'{n}ski \& Miko{\l}ajewska 1998).  This accretion disc is fully specified with confidently measured parameters, except that the accretion rate is a free parameter adjusted to fit the ultraviolet brightness.  The X-ray flux measures are not included in the fit, as they come from a different physical mechanism (boundary layer emission) that is not described by the $\alpha$-disc model.

My fitted temperature for the red giant is 2870$\pm$40 K, with this being applicable both for the low-state and the high-state.  The fitted disc model (the light-blue curve) is essentially determined to fit the ultraviolet fluxes.  The vertical scatter in the SED plot is due to the intrinsic variability of T CrB, with this variability being large in the ultraviolet.  In the high-state, the ultraviolet flux is larger than in the low-state by typically a factor of 20$\times$, and this points to the high-state having a greatly increased accretion rate.

\begin{table}
	\centering
	\caption{T CrB Spectral Energy Distribution (full table of 55 fluxes, plus column for references, appears in the on-line Supplementary Material}
	\begin{tabular}{llllll}
		\hline
		Source &  Year  &    Band & Flux   &  Log[$\nu$] &   Log[$F_{\nu}$] \\
		\hline
		
Walker	&	1976	&	{\it U}	&	12.85 mag	&	14.92	&	-1.76	\\
Walker	&	1976	&	{\it B}	&	11.6 mag	&	14.83	&	-0.91	\\
Walker	&	1976	&	{\it V}	&	10.04 mag	&	14.74	&	-0.37	\\
Walker	&	1976	&	{\it R}	&	9.04 mag	&	14.67	&	-0.07	\\
Walker	&	1976	&	{\it I}	&	7.46 mag	&	14.58	&	0.46	\\
...	&		&		&		&	&    \\
AID  	&	2016--22	&	{\it I}	&	7.4 mag	&	14.58	&	0.49	\\
{\it Swift} XRT	&	2018	&	0.2--10 keV	&	5.6$\times$10$^{-6}$ Jy	&	17.38	&	-5.25	\\
{\it Swift} UVOT 	&	2019--22	&	{\it UVW1}	&	12.88 mag	&	15.08	&	-2.09	\\
{\it Swift} UVOT 	&	2019--23	&	{\it UVM2}	&	10.72 mag	&	15.13	&	-1.20	\\
{\it Swift} UVOT 	&	2019--24	&	{\it UVW2}	&	10.98 mag	&	15.18	&	-1.32	\\

		\hline
	\end{tabular}		
\end{table}

\begin{figure}
	\includegraphics[width=1.01\columnwidth]{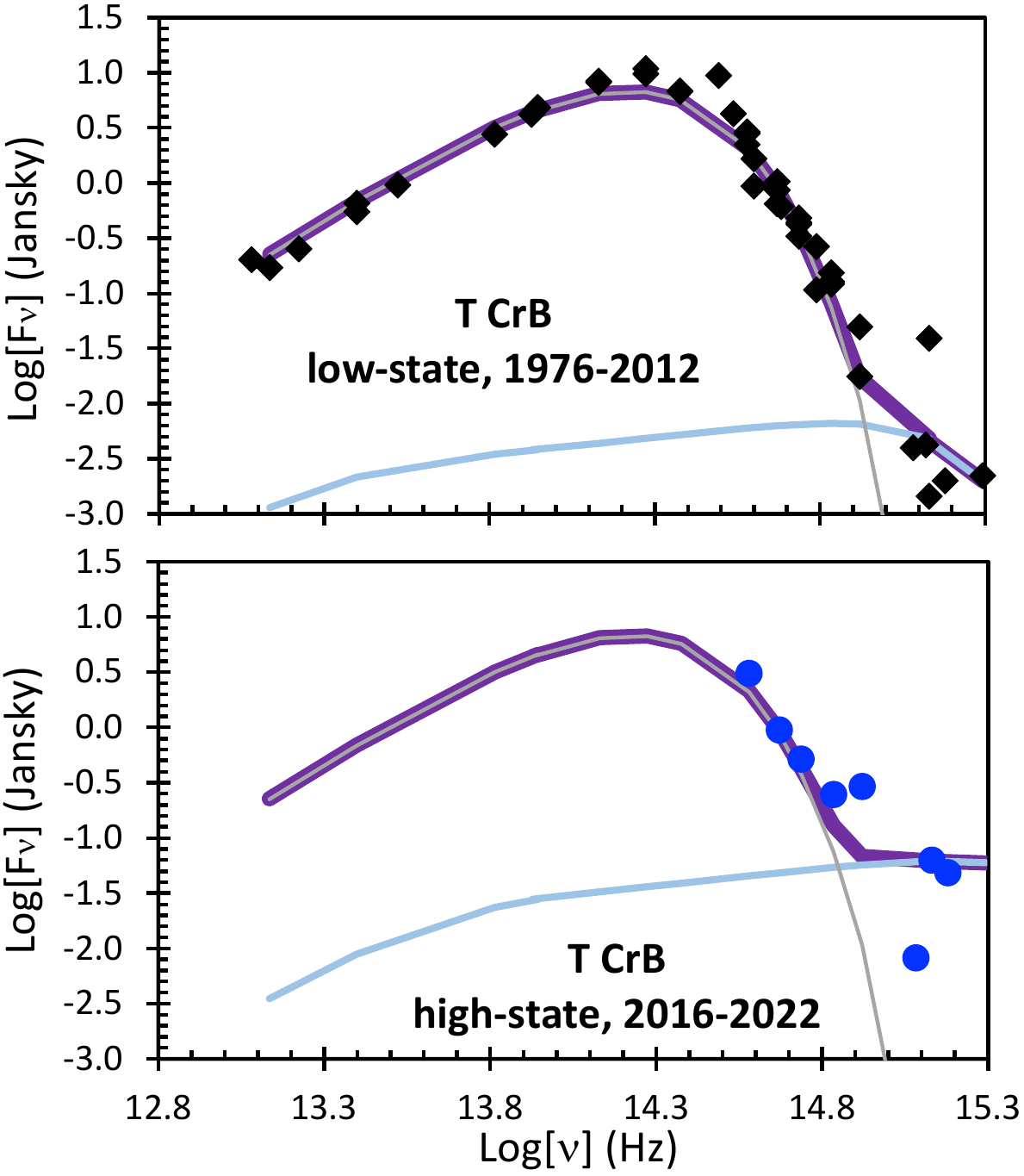}
    \caption{Spectral energy distribution plot for quiescent low-state (top panel) and the recent high-state (bottom panel).  The fitted blackbody from the red giant companion star is shown by the thin gray line, and this is unchanged from low-state to high-state.  The best-fitting temperature is 2870$\pm$40 K.  The fitted $\alpha$-disc is shown by the light-blue curve, with the position of the shape only determined from the ultraviolet fluxes.  Redward of the $B$-band, the red giant light dominates over the disc light by orders-of-magnitude.  The scatter about the best smooth curve shows the ordinary temporal flickering of T CrB.  Top panel:  With many flux measures from 1976--2012, the SED is well defined, with a good fit to the red giant's blackbody, while the highly-variable disc contribution is only visible in the ultraviolet.  Bottom panel:  I can find few useable measures of the flux after the high-state turn-on around 2015.  These show that the $R$ and $I$ magnitudes are unchanged from low- to high-state, which is to say that the red giant is not the source of the extra luminosity in the high-state.  The extra light in the high-state is very blue and ultraviolet rich, and this light varies substantially day-to-day.}  
\end{figure}

\section{ACCRETION RATE AND ACCRETION MASS}

The activity and energetics and eruptions of T CrB are all driven by the accretion via Roche lobe overflow from the red giant to a disc around the white dwarf.  The critical parameters are the accretion rate ($\dot{M}$) and the mass accumulated on the surface of the white dwarf ($M_{\rm acc}$).
The accretion light is not readily separable from the red giant light in the optical bands, while the accretion light dominates in the ultraviolet.  So $\dot{M}$ can only be measured from the ultraviolet flux.  Selvelli et al. (1992) measured the accretion at 2.32$\times$10$^{-8}$ M$_{\odot}$ yr$^{-1}$, based on many measures with the {\it International Ultraviolet Explorer} ({\it IUE}).  In 25 observations from 1979--1990, they found the ultraviolet flux to be highly variable, with the RMS being 60 per cent of the average, and with an extreme range of a factor of 20$\times$.  With this variability, we can only derive an average accretion rate that is hopefully representative of the entire time interval.  The tool is the SED fits to the ultraviolet flux, as in the previous section, where the $\alpha$-disc model is completely determined with good accuracy, except for the accretion rate, which can be determined as a fit parameter.

For the low-state, we have 25 ultraviolet fluxes from {\it IUE} in 1979--1990, 2 fluxes from {\it Galex} in 2005 and 2007, and15 fluxes in three bands from {\it Swift} UVOT in 2008 and 2011  For the high-state, we have 8 fluxes in three bands from {\it Swift} UVOT in 2019--2022, plus 4 $U$ magnitudes reported by AID for 2021 and 2022.  The fitted $\alpha$-disc models are shown in Fig. 14, with the derived $\dot{M}$ representing the average over the time intervals.  The average low-state accretion rate is 3.2$\times$10$^{-9}$ M$_{\odot}$ yr$^{-1}$, while it is 6.4$\times$10$^{-8}$ M$_{\odot}$ yr$^{-1}$ in the high-state.

T CrB has gone through many states since the first positive detection in 1855.  In Table 8, I have listed the states, with a descriptive name and the observed range of years.  To each of these states, I have assigned either the low-state or the high-state accretion rate, with the eruptive states presumably having no accretion as the overflow mass is blown away.  The mass accreted in each interval, $M_{\rm acc}$ in solar masses, is then calculated and listed.  With this, the total mass accreted between eruption, $\Sigma$$M_{\rm acc}$ can be added up.  The 1866--1946 interval, one complete eruption cycle had 1.4$\times$10$^{-6}$ M$_{\odot}$.  The 1946--2022.8 interval has an accreted mass of 1.2$\times$10$^{-6}$ M$_{\odot}$, but this cycle is not yet complete.

My value for $\log$[$\Sigma$M$_{acc}$] for one complete eruption cycle is $-$5.9, and this can be compared to theoretical predictions.  The mass accreted between eruptions required to trigger the nova eruption is variously labelled as $M_{\rm ign}$ by Shen \& Bildsten (2009), $m_{\rm acc}$ by Yaron et al. (2005), or $M_{\rm trigger}$ by Schaefer et al. (2022a).  These calculations are for a steady-state accretion, yet for T CrB it is unclear whether the low-state or high-state accretion is applicable.  In their fig. 7, Shen \& Bildsten give the trigger mass for a 1.35 M$_{\odot}$ white dwarf being fed material with solar abundances, with the log-mass ranging from $-$5.9 to $-5.2$ for $\log[\dot{M}]$ varying from $-$7 to $-$9.  Similarly, for varying accretion rate over the same range, Yaron et al. have ranges of log-mass from $-$7.1 to $-$6.4 for relatively cool white dwarfs, and from $-$7.1 to $-$6.7 for relatively hot white dwarfs.  I can only conclude that theory has not yet made a confident and accurate prediction, with the log of the trigger mass possibly being from $-$5.2 to $-$7.1.  With this, my measure fits in well with theoretical expectations for the trigger mass.

\begin{table*}
	\centering
	\caption{T CrB Accretion Rates and Energetics}
	\begin{tabular}{lllllll}
		\hline
		State &  Years  &    $\dot{M}$ & $M_{\rm acc}$   &  Log[$E_{\rm B}$] &   Log[$E_{\rm V}$]   &   Log[$E_{\rm bolo}$] \\
		\hline
		
Primary eruption	&	1866.361	 --	1866.44	&	0	&	0	&	...	&	43.14	&	44.20	\\
Inter-eruption high-state	&	1866.44	 --	1866.66	&	6.4$\times$10$^{-8}$	&	1.4$\times$10$^{-8}$	&	...	&	41.20	&	42.76	\\
Secondary eruption	&	1866.66	 --	1866.95	&	0	&	0	&	...	&	42.17	&	43.23	\\
Post-eruption high-state	&	1866.95	 --	1874.8	&	6.4$\times$10$^{-8}$	&	5.0$\times$10$^{-7}$	&	...	&	42.63	&	44.19	\\
Low-state	&	1874.8	 --	1935.5	&	3.2$\times$10$^{-9}$	&	1.9$\times$10$^{-7}$	&	42.29	&	42.28	&	43.58	\\
Pre-eruption high-state	&	1935.5	 --	1942.5	&	6.4$\times$10$^{-8}$	&	4.5$\times$10$^{-7}$	&	42.78	&	41.90	&	43.46	\\
Pre-eruption dip	&	1942.5	 --	1946.103	&	6.4$\times$10$^{-8}$	&	2.3$\times$10$^{-7}$	&	42.49	&	41.41	&	42.97	\\
Primary eruption	&	1946.103	 --	1946.19	&	0	&	0	&	43.72	&	43.11	&	44.17	\\
Inter-eruption high-state	&	1946.19	 --	1946.38	&	6.4$\times$10$^{-8}$	&	1.2$\times$10$^{-8}$	&	41.28	&	41.02	&	42.59	\\
Secondary eruption	&	1946.38	 --	1946.68	&	0	&	0	&	42.37	&	42.09	&	43.15	\\
Post-eruption high-state	&	1946.68	 --	1954.5	&	6.4$\times$10$^{-8}$	&	5.0$\times$10$^{-7}$	&	42.80	&	42.46	&	44.03	\\
Low-state	&	1954.5	 --	2015.0	&	3.2$\times$10$^{-9}$	&	1.9$\times$10$^{-7}$	&	42.29	&	42.28	&	43.58	\\
Pre-eruption high-state	&	2015.0	 --	2022.8	&	6.4$\times$10$^{-8}$	&	5.0$\times$10$^{-8}$	&	42.76	&	42.59	&	44.16	\\

		\hline
	\end{tabular}		
\end{table*}

\section{ENERGETICS OF ERUPTIONS AND HIGH-STATES}

The total energy radiated by T CrB in its various states can tell us about the physics of the states.  This energy can be calculated from integrals under the long-term light curves, to produce the energy in the $B$ and $V$ bands, $E_{\rm B}$ and $E_{\rm V}$.

The first step is to pull out binned $B$ and $V$ light curves.  For this, I use the light curve shown in Fig. 6, where the ellipsoidal variations are averaged out by the use of 113.7843-day bins during quiescence.  

The second step is to correct for extinction.  With an adopted $E(B-V)$=0.065 from {\it Galex}, the $B$-band and $V$-band extinctions are 0.27 and 0.21 mag.

The third step is to convert all the magnitudes to flux densities with units of Jansky.  For this, a zero-magnitude star has flux densities of 4260 and 3640 Jy in the $B$ and $V$ bands (Bessell 1979).

The fourth step is to subtract out the constant light of the red giant, so we are left with only flux from the accretion processes.  The subtracted red giant fluxes in $B$ and $V$ are taken from the SED fits.

The fifth step is to convert these flux densities in Janskys to the luminosity in the $B$ and $V$ bands, with units ergs per second.  Three conversion factors are needed.  First, the flux density in Janskys is to be multiplied by a factor of $10^{-23}$ to get units of erg s$^{-1}$ cm$^{-2}$ Hz$^{-1}$.  Second, the factor of $4\pi D^2$ is used to get units of erg s$^{-1}$ Hz$^{-1}$.  Here, I adopt the distance to T CrB of 914 pc (Schaefer 2022c).  Third, the converted flux must be multiplied by the bandwidth, in Hertz, to get the luminosity in that band.  For bandwidths of 980~\AA ~and 890~\AA ~in the $B$ and $V$, the correction factors are 1.54$\times$10$^{14}$ and 8.88$\times$10$^{13}$ Hz.  

The sixth step is to integrate the luminosity over time for various intervals so as to get the total radiated energy $E_{\rm B}$ and $E_{\rm V}$ (Table 8).  With the pre-eruption dip apparently caused by dust extinction from the circumstellar medium, the observed brightness does not represent the emitted energy, so I have calculated the radiated energy assuming the high-state luminosity.

The seventh step is to correct from $E_{\rm B}$ and $E_{\rm V}$ to the bolometric energy $E_{\rm bolo}$.  To make this correction for the low- and high-states, I use my SED fits for the $\alpha$-disc models.  For the primary and secondary eruptions, I adopted an SED for a 10,000 K blackbody, as appropriate for the photosphere of nova shells.  My bolometric range extends from 0.1--30 microns, which includes effectively all the energy, except for the unknown flux that comes out in the far-ultraviolet.  The X-ray flux is always greatly too small to contribute any significant fraction of energy.  Both the $B$ and $V$ light curves should produce the same $E_{\rm bolo}$, to within measurement errors.  The differences in the two measures of $\log[E_{\rm bolo}]$ have an average near zero and an RMS of 0.2, which I take to be the real measurement uncertainty.  The $\log{E_{\rm B}}$, $\log{E_{\rm V}}$, and $\log[E_{\rm bolo}]$ values are in Table 8.

\section{THE UNIQUE STATES AROUND ERUPTIONS}

\subsection{The High-State}

The unique high-state starts close to ten years before the 1946 eruption (and $<$10.1 years before the 1866 eruption), continuing until nine years after both eruptions as a nearly flat plateau, for a duration of 19 years.  (Superposed on this high-state plateau are the pre-eruption dip, the primary eruption, and the secondary eruption.)  No precedent is known.  Back in the late 1930's, it was recognized that the extra light was very blue in colour, and powering the new bright emission lines for high-ionization species (Payne-Gaposchkin \& Wright, 1946).  So the situation for the high-state was recognized to be some mysterious source that was very hot.

The appearance of the hot source is apparently associated with mass ejection at some level.  During World War II, spectroscopists started recording P Cygni line profiles indicating associated outflows at $\sim$300 km s$^{-1}$ (Swings, Elvey, \& Struve 1940).

As for the mechanism and cause of this hot-source, I am aware of no suggestions in the literature, largely because the existence of the high-state was unknown until the long-term fully-calibrated light curve (as in Fig. 1 and Fig. 6) became available (Schaefer 2014).  Nevertheless, there are only two possibilities, either the accretion rate suddenly increased by roughly 20$\times$ making for a bright accretion disc, or some type of nuclear burning ignited on the surface of the white dwarf powering some sort of a hot photosphere.  The nuclear burning possibility is dubious because there is no situation where burning would simmer steadily for 19 years, much less punctuated by at least one real thermonuclear runaway on the white dwarf.  Further, there is no real chance that a hot photosphere around the white dwarf would drive mass ejection with a velocity as slow as 300 km s$^{-1}$.  

So the high-state is an accretion event, where the red giant suddenly starts (and stops 19 years later) to pour mass through the Roche lobe with the $\dot{M}$ value 20$\times$ higher.  I know of no physical mechanism that can drive this accretion high-state.  Further, I know of no timing mechanism that can start the high-state at a time 69$\pm$1 year after the prior eruption, that starts the mass ejection several years before the eruption, and stops the high-state 9$\pm$1 years after the eruption.

The existence of the pre-eruption high-state presents a mystery as to why the system brightens {\it before} the eruption.  The physical mechanism that makes the high-state is presumably associated with the accretion rate, which is controlled by the Roche lobe overflow in the atmosphere of the red giant near the inner Lagrangian point of the orbit.  The physical mechanism that determines the timing of the start of the classical nova eruption is controlled by the pressure and temperature at the base of the accreted material on the surface of the white dwarf.  But how can the atmosphere of the red giant star know about the far-future conditions deep in the white dwarf?  What mechanism connects the two locations?  One suggestion is that the red giant atmosphere independently kicks into a high-state, and the enlarged accretion rate quickly drives the white dwarf accretion layer to the nova trigger threshold.  But this is a description of the problem (that the high-state starts 10 years before the nova event), not an explanation for the timing of the high-state and the eruption.  So we are left with no explanation for the enigma as to why the high-state anticipates the eruption by 10 years, just as we have no explanation for the cause or physical mechanism of the high-state existence.

Intriguingly, I know of three other novae that have conspicuous pre-eruption rises that might be related to the pre-eruption high-state of T CrB:  {\bf (1)} V533 Her had a distinct exponential rise by 1.5 mag in the 1.5 years leading up to its 1963 nova eruption (Collazzi et al. 2009).  {\bf (2)} V1500 Cyg had an exponential rise in brightness by $>$8 mags in the 23 days before its eruption (Collazzi et al. 2009).  {\bf (3)} The 2011 eruption of RN T Pyx had a unique flare that peaked at 1.0 mag above the normal quiescent level at a time 9 days before the fast rise of the nova event (Schaefer et al. 2013).  

Further, 8 novae are known to have a prolonged high-state lasting for many decades after the eruption has completely stopped (Schaefer \& Collazzi 2010).  These are labelled as `V1500 Cyg stars', named after their prototype.  All of these classical novae are more than 2.5 mag brighter than their pre-eruption levels for times more than thirty years after the nova light curve has gone essentially flat.  The only plausible explanation is that the white dwarf has continued nuclear burning on its surface, and this is driving a high accretion rate.  The cause and mechanism for this continued nuclear burning is unknown.  These post-eruption high-states might serve as precedence or exemplars for the T CrB post-eruption high-state.

We are left with the T CrB pre- and post-eruption high-state as being completely unique, yet there are at least 11 nova systems with pre- and post-eruption high-states that might share the same physical mechanism.  Further, three other nova systems have re-brightenings that can be labelled as secondary eruptions (see Section 11.3).  All of these novae have energetic transient high-states and eruptions that are mixed together with a wide array of morphologies and time-scales.  The physical mechanisms are not proven in any of these other cases, but it appears that all share a greatly increased accretion rate, perhaps powered by continued nuclear burning on the white dwarf that starts and stops away from the regular nova event.  With these loose precedents, I expect that the high-state of T CrB is the result of a greatly increased accretion rate, possibly triggered by steady thermonuclear burning on the white dwarf.  

\subsection{Pre-Eruption Dip}

There is no precedent in any system for a pre-eruption dip.  The primary clue as to the cause of the dip is that the $V$-band light falls roughly 1.5 mag fainter than the normal quiescent level of the red giant alone.  I think that the easiest mechanism to dim the red giant light, to far fainter than its normal level for most of a year, is for the dimming to be external to the star.  With this, the fainter brightness of the red giant is likely due to dimming by circumstellar dust.  This fits nicely with the observed P Cygni profiles spotted only in the several years before the eruption (Payne-Gaposchkin \& Wright 1946).  That is, just before the start of the pre-eruption dip, T CrB was ejecting gas with velocities $\sim$300 km s$^{-1}$, while ejecta from novae are famous for occasionally producing shells of freshly-made dust that dims the star, often by many magnitudes.  So we have a simple explanation for the pre-eruption dip.

Details still need to be worked out.  For example, models of dust formation can test whether the timing between the visibility of the P Cygni profiles and the dimming light curve is reasonable, and ejection rates should be calculated.  Maybe most important is to understand the relative dimming between $B$ and $V$, which the larger dimming in the $V$ appears in violation of the usual extinction laws.  Fortunately, in the next year or so, T CrB might go through another pre-eruption dip, and this can be observed with the full suite of modern instruments.

\subsection{The Secondary Eruption}

The secondary eruption behaved identically after the 1866 and 1946 eruptions (see Fig. 3).  In both cases, the nova light had completely stopped 30 days after the peak, returning to the high-state level, only to have the secondary eruption start around day 110, have a nearly-flat maximum peaking around day 160, and then suddenly drop to the high-state level around day 210.

The secondary eruption displayed a continuum spectrum, indicating some sort of an optically-thick emission region.  The $B-V$ is the same as for the colour at the primary peak, suggesting that the optically-thick region is some sort of a photosphere with a 10,000 K temperature, as is universal for nova eruptions.  The total bolometric energy of the secondary eruption is 10$\times$ less than that of the primary eruption.  This total energy (10$^{43.2}$ ergs) is large, similar to the energy from thermonuclear runaways of nova events, and greatly larger than anything possible from accretion energy.

I know of three attempted theoretical explanations:  {\bf (1)} The first attempt appeared in the highly influential review on RNe by Webbink et al. (1987), with the secondary eruption caused by a concentrated ring of gas in the disc suddenly accreting on to the surface of the main-sequence star.  This model requires that the white dwarf be replaced by a main-sequence star, with this idea being strongly refuted (Selvelli et al. 1992).  {\bf (2)} Webbink et al. (1987) mention a second possible explanation for the secondary maximum, and that has the inner hemisphere on the red giant being irradiated by the nova light, with this reprocessed energy becoming visible as the hot/bright side of the companion star orbiting into and then out of view.  This possibility is ruled out by the start of the secondary eruption being 80 days after the primary eruption light had completely faded away.  Further, Webbink et al. (1987) points out that the timing and peak colours are all wrong for the irradiation explanation.  {\bf (3)} Hachisu \& Kato (1999) proposed an extension to the irradiated companion idea, where a severely tilted disc is irradiated, with the combined irradiation resulting in the secondary event.  This idea fails to all the same problems as the previous explanation.  Further, the irradiation of the companion star and of the accretion disc cannot produce a spectrum that shows a continuum with a photospheric temperature of around 10,000 K.  The effects of the disc tilting are much too small to matter, as disc flux scales as the cosine of the viewing angle, with the postulated tilt going from 70$\degr$ to 35$\degr$, for a 2.4$\times$ increase in the disc light, which is only a few per cent of the optical light.  In the end, the three attempted explanations all fail.

Let me propose a new explanation.  The idea is simply that the secondary eruption is a separate nova eruption, i.e., a thermonuclear runaway involving accreted mass on the surface of the white dwarf.  The nuclear burning is of mass accreted at the high-$\dot{M}$ rate between the end of the primary nova eruption and the start of the secondary nova eruption.  The trigger mass is greatly smaller for the secondary eruption because the surface of the white dwarf is greatly hotter due to the primary eruption.  This idea provides a simple explanation for the continuum from a 10,000 K photosphere as being the usual consequence of a thermonuclear nova event.  This idea provides a known energy source that is sufficiently large to explain the observed radiative energy of 10$^{43.2}$ ergs.  The fast rise and the duration are characteristic of nova eruptions, although the light curve shape is more like the uncommon low-energy F-class novae (Strope et al. 2010).  The relative delay from the primary to secondary eruptions is determined by the high-state accretion rate, which appears constant between the 1866 and 1946 eruptions.  However, before this idea can be taken seriously, detailed model calculations of nova eruptions are needed for the specific conditions of T CrB.  Critical questions would be whether the different conditions at the start of the secondary eruption could lead to the observed light curve shape, and whether the accreted mass before the secondary eruption is adequate to produce the observed energy.

Intriguingly, some rare classical novae have low-amplitude eruption events that are soon after their primary nova:  {\bf (1)} V1047 Cen had a pre-eruption magnitude $V$=18.7, came to a peak of $V$=8.5 in mid-2005, and slowly faded to 16th mag from 2006--2019.  In mid-2019, the system brightened suddenly by $>$4 mags, staying with a flat-topped peak around 14.0 mag for 10 months, and then faded back to its pre-outburst level.  {\bf (2)} V5856 Sgr had a pre-eruption brightness of $I$$>$22, a peak magnitude of $V$=5.9 in 2016, and a slow decline from 11.0--13.3 mag over the interval 2017.0--2021.6.  Then, the old nova suddenly brightened by over 1 mag for a flat-topped re-brightening lasting 260 days, followed by three 2-month-long flares that are continuing to the present.   {\bf (3)} V1280 Sco has a pre-eruption level of $V$>20.0, reached a peak at $V$=3.78 in 2007, had the usual D-class jitters, went into a deep dip fading below 16.0 mag (likely due to the usual dust formation in the expanding nova shell), and the light curve recovered to 11.5 mag in early 2008.  In all other cases, D-class novae will immediately start fading, but V1280 Sco remained constant at around 11.5 mag from 2008--present.  This long luminous plateau might be a high-state, or it might be a second eruption starting just 1 year after the primary eruption with a low flat-topped light curve.

So the secondary eruption of T CrB might have partial precedent with these three other novae.  The morphologies and time-scales of all four secondary eruptions vary widely, but they all share the properties of a large total energy output (comparable or larger than for the primary eruption), starting out with the nova brightness (and hence accretion) far above its pre-eruption level, having the start of the secondary eruption sufficiently close in time to the primary eruption (from 0.3--14 years) such that some causal connection seems forced, with all the secondary eruptions having fast rises of under a month followed by a flat-topped plateau.  These morphological similarities are not enough to prove that the same mechanism is powering all four novae, but it is enough to make a reasonable case that we are seeing four versions of one eruption mechanism.  The four novae are a fair cross-section of the nova population, and I do not see any pattern to the four novae with secondary eruptions.

\section{ACKNOWLEDGEMENTS}

The real heroes of this massive data-mining program are the roughly-2000 observers from 1842--2022, who spent many nights in vigils over T CrB.  Further, this heartfelt thanks must extend to the roughly 150 workers involved with archiving and maintaining the plates, letters, data books, light curves, and papers throughout the last 180 years.  Ron Webbink (University of Illinois, Champagne-Urbana) provided the valuable service of creating an exhaustive bibliography on T CrB for papers, notebooks, letters, and archival manuscripts, with these resources dominating the pre-1975 visual and photographic light curves, with this preventing many of the old magnitudes from being lost.  Further, many instrument builders, observers, and data-handlers provided light curves from APASS, DASCH, {\it TESS}, and ASAS.  For my study, I made heavy use of many utilities and products of the AAVSO, including comparison star calibrations (APASS), finder chart construction (VSP), light curve plotting (LCG), Fourier transforms (VStar), archiving over one-eighth of a million T CrB magnitudes from observers worldwide (AID), plus the old charts, manuscripts, and letters now only in the AAVSO archives and files.  I have been a heavy user of the Harvard plates, with the courtesy of Josh Grindlay and Alison Doane.  I thank Peter Kroll for hospitality during two long visits to Sonneberg Observatory, and Ulrich Heber for hospitality during a visit to Bamberg Observatory.  The referee was helpful with the {\it TESS} analysis.

The DASCH project at Harvard is grateful for partial support from NSF grants AST-0407380, AST-0909073, and AST-1313370

\section{DATA AVAILABILITY}

All of the photometry data are explicitly given in Tables 2--6, or are publicly available from the references and links in Table 1.


{}

%%%%%%%%%%%%%%%%%%%%%%%%%%%%%%%%%%%%%%%%%%%%%%%%%%
% Don't change these lines
\bsp	% typesetting comment
\label{lastpage}
\end{document}